\documentclass[nonacm]{acmart}
\usepackage{pgfplots}
\usetikzlibrary{patterns,,patterns.meta}
\usepackage{pgfplotstable}
\usepackage{algorithm,algpseudocode}
\usepackage{booktabs, multirow}
\usepackage{makecell}
\usepackage{graphicx} 
\usepackage{array}
\usepackage{amsmath}
\usepackage{tabularx}
\usepackage{subcaption} 
\usepackage{float}
\pgfplotsset{compat=1.18} 
\usepackage{tikz}
\usepackage{url}
\usepackage{xcolor}

\definecolor{oiBlue}{RGB}{0,114,178}
\definecolor{oiOrange}{RGB}{230,159,0}
\definecolor{oiSkyBlue}{RGB}{86,180,233}
\definecolor{oiGreen}{RGB}{0,158,115}
\definecolor{oiYellow}{RGB}{240,228,66}
\definecolor{oiRed}{RGB}{213,94,0}
\definecolor{oiPink}{RGB}{204,121,167}
\definecolor{myGrey}{RGB}{179,179,179}

\AtBeginDocument{%
  }

\newcommand{\mat}[1]{\mathbf{#1}}
\newcommand{\vect}[1]{\mathbf{#1}}
\newcommand{\erdosrenyi}{Erd\H os-R\'{e}nyi}
\newcommand{\transpose}     {^{\mbox{\scriptsize \sf T}}}

\newcommand{\nnz}{\textit{nnz}}

\newcommand{\nthreads}{\texttt{nthreads}}
\newcommand{\SendBuffer}{\texttt{SendBuffer}}

\begin{document}

\pgfplotstableread[col sep=comma]{
procs,com-Orkut+CB,wikipedia-2007+CB,soc-LiveJournal1+CB,wb-edu+CB,uk2002+CB,europe-osm+CB,twitter7+CB,Gap-road+CB,com-Orkut+ours,wikipedia-2007+ours,soc-LiveJournal1+ours,wb-edu+ours,uk2002+ours,europe-osm+ours,twitter7+ours,Gap-road+ours
1,1189.75,248.488,365.95,331.5,1579.56,793.64,8148.86,425.12,40.09,8.45,12.76,13.02,56.77,32.01,269.89,17.87
4,541.834,117.4,225,155.433,705.53,284.9,3694.84,149.189,11.2,2.642,4.526,4.085,17.2,13.108,69.07,7.28
16,188.32,60.22,124.43,56.92,351.98,111.6,1744.77,57.07,4.59,1.41,2.8,1.84,8.16,5.13,39.53,2.68
64,80.21,31,56.56,27.68,175.16,46.41,831.49,24.24,1.82,0.66,1.26,0.899,3.68,2.05,18.48,0.99
256,44.09,16.3,23.87,15.67,87.16,24.33,319.16,,0.97,0.36,0.59,0.43,1.73,0.93,6.99,
1024,16.34,7.79,11.253,8.633,42.35,13.649,139.31,,0.379,0.182,0.302,0.23,0.95,0.48,2.94,
}\scalingdatatable

\pgfplotstableread[col sep=comma]{
procs,GAP-web+CB,com-Friendster+CB,webbase+CB,GAP-urand+CB,GAP-web+ours,com-Friendster+ours,webbase+ours,GAP-urand+ours
16,2181.33,,1369.42,,49.27,,30.15,
64,1008.9,1511.45,663.2,474.4,20.9,32.4,13.62,19.97
256,542.48,534.94,342.58,137.8,9.94,11.3,6.26,6.46
1024,383.505,207.17,179.65,45.95,6.213,4.11,3.3,1.94
}\scalingdatalargetable

\pgfplotstableread[col sep=comma]{
nprocs,com-Orkut+cB,wikipedia-2007+cB,soc-LiveJournal1+cB,wb-edu+cB,uk-2002+cB,Gap-road+cB,twitter7+cB,europe_osm+cB,GAP-web+cB,webbase+cB,com-Friendster+cB,GAP-urand+cB,com-Orkut+ours,wikipedia-2007+ours,soc-LiveJournal1+ours,wb-edu+ours,uk-2002+ours,Gap-road+ours,twitter7+ours,europe_osm+ours,GAP-web+ours,webbase+ours,com-Friendster+ours,GAP-urand+ours
4,6.62,3.87,3.16,0.95,4.71,1.39,58.32,2.44,,,,,0.342,0.079,0.131,0.105,0.456,0.12,2.2,0.237,,,,
16,4.4,3.03,1.97,1.36,5.81,1.92,28.95,3.34,47.46,19.07,,,0.16,0.047,0.09,0.052,0.21,0.084,1.18,0.15,1.18,0.76,,
64,3.3,1.7,1.52,1.78,6.89,2.16,18.5,4.07,57.19,22.4,43.07,37.35,0.09,0.03,0.06,0.03,0.12,0.05,1.12,0.12,0.69,0.52,1.29,1.27
256,2.6,1.34,1.28,1.9,7.98,,12.97,4.48,71.03,25.08,30.35,27.4,0.05,0.02,0.037,0.02,0.07,,0.53,0.072,0.36,0.27,0.67,0.418
1024,2.29,1.12,1.13,2.51,8.84,,9.85,4.7,79.98,28.81,22.66,18.83,0.029,0.013,0.023,0.023,0.058,,0.274,0.048,0.23,0.16,0.34,0.159
}\scalingcommunication

\pgfplotstableread[col sep=comma]{
Threads,uk-2002,twitter7,GAP-web,GAP-urand
1,6.447,52.55,38.4,28.88
2,3.52,27.28,21.03,15.47
4,2.14,14.81,12.98,8.23
8,1.5,9.46,8.52,5.29
16,1.145,6.65,6.33,3.87
32,0.974,5.278,5.513,2.99
}\threadscalingdatatable

\pgfplotstableread[col sep=comma]{
Threads,uk-2002,twitter7,GAP-web,GAP-urand
1,6.447,52.55,38.4,28.88
2,3.52,27.28,21.03,15.47
4,2.14,14.81,12.98,8.23
8,1.5,9.46,8.52,5.29
16,1.145,6.65,6.33,3.87
32,0.974,5.278,5.513,2.99
}\reorderingbreakdown

\pgfplotstableread[col sep=comma]{
procs,wikipedia-2007+CB,soc-LiveJournal1+CB,twitter7+CB,com-Friendster+CB,GAP-urand+CB,nlpkkt+CB,wikipedia-2007+ours,soc-LiveJournal1+ours,twitter7+ours,com-Friendster+ours,GAP-urand+ours,nlpkkt+ours
1,475.069,485.25,,,,,14.39,14.63,,,,
4,232.9,307.9,,,,1848.43,10.317,9.8,,,,51.93
16,133.94,184.4,,,,932.9,5.7,5.4,263.62,,,25.39
64,76.43,97.58,1722.4,1632.4,477.9,458.19,2.9,2.62,148.25,158.16,22.95,12.918
256,39.4,39.6,756.8,680.3,135.5,236.99,1.31,1.33,93.7,72.87,8.54,6.93
1024,20.14,18.32,321.1,248.8,45.59,129.981,4.1,4.2,36.3,31.1,3.52,6
}\scalingReordering

\pgfplotstableread[col sep=comma]{
procs,our-twitter7,our-friendster,our-webbase,our-gapurand,our-metaclust,CB-twitter7,CB-friendster,CB-webbase,CB-gapurand,CB-metaclust
1,,,,,,12264.78,,7442.5,,
4,16.1,,14.29,,,5206.81,,4529.4,,
16,8.56,16.38,6.59,8.56,,2974.33,5925.4,2366.26,2100.91,
64,4.32,7.17,3.22,4.33,,1484.4,2662.7,1320.3,983.07,
256,3.38,4.33,3.19,2.25,4.48,584.33,650,661.22,526.8,
1024,9.5,8.84,5.9,8.273,1.27,254.5,348.04,332.8,43.78,430.35
}\scalingExtract

\pgfplotstableread[col sep=comma]{
procs,our-twitter7,our-friendster,our-webbase,our-gapurand,our-metaclust,CB-twitter7,CB-friendster,CB-webbase,CB-gapurand,CB-metaclust
1,,,,,,,,,,
4,111.45,,,,,5536,,,,
16,41.33,89.46,22.43,36.14,,2975.26,,2268.3,2389.53,
64,15.9,30.68,9.37,13.69,,1519.88,2711.08,1282.21,605.6,
256,6.71,10.51,5.77,5.2,,591,992.7,604.3,158.7,
1024,10.36,11.07,10.1,9,7.25,257.48,352.6,335.92,52.42,498.25
}\scalingExtractt
\pgfplotstableread[col sep=comma]{
procs,our-Orkut,our-wbedu,our-twitter,our-friendster,our-metaclust,CB-orkut,CB-wbedu,CB-twitter,CB-friendster,CB-metaclust
1,26.8,7.11,215.08,,,5839.91,1385.87,,,
4,14.12,3.72,99.83,,,2715.14,775.22,,,
16,5.15,1.92,52.53,99.89,,1059.27,344.94,,,
64,3.01,1.48,27.4,46.01,,473.6,184.69,4992.89,,
256,4.12,4.71,13.78,9.88,,277.85,100.41,1979.29,1979.29,
1024,14.29,10.79,15.28,16.53,,106.85,59.98,852.32,1185.32,
}\scalingAsgn

\pgfplotstableread[col sep=comma]{
procs,our-Virus,our-Eukrya,our-Archae,CB-Virus,CB-Eukrya,CB-Archae
1,0.962317,55.3477575,23.11971889,257.082788,25006.88388,10011.79361
4,0.344531,14.162961,6.094343,61.371689,6122.689969,2465.631074
16,0.814774,4.180342,2.382777,15.020824,1460.799406,618.005063
64,2.699544,3.372775,3.040641,5.460667,356.191551,172.035173
256,9.40193,10.738854,11.61809444,13.52245,98.888221,60.13896444
}\scalingExtClust

\pgfplotstableread[col sep=comma]{
procs,MetaCluste-ours,hyperlink-ours,MetaCluste-CB,hyperlink-CB
256,36.820689,,346.046506,
1024,9.908137,53.129008,164.21749,
4096,3.838198,21.561657,84.850811,
}\scalingLarge

\pgfplotstableread[col sep=comma]{
Dataset,node,nnz,Directed,avgDeg,GraphBLAS,OurPerm1,OurPerm2,OurPerm3,Petsc
com-Orkut,3M,234M,Undirected,76,2.04,9.3,5.5,4.11,22.7
wikipedia-2007,3.5M,45M,Directed,12,0.27,2.18,1.5,1.32,5.43
soc-LiveJournal1,4.8M,68.9M,Directed,14,0.47,3.7,2.9,2.32,8.36
wb-edu,9.8M,57M,Directed,5,0.49,3.31,2.3,1.82,9.84
uk-2002,18.5M,298M,Directed,16,1.4,13.7,9.9,7.63,34.73
Gap-road,23.9M,57M,Undirected,1,1.1,5.2,3.8,2.64,17.22
europe-osm,50.6M,108M,Undirected,2,1.6,10,8.3,6.05,35.7
}\PermGraphBLASOurs

\title{Distributed-memory Algorithms for Sparse Matrix Permutation, Extraction, and Assignment}

\author{Elaheh Hassani}
\affiliation{%
  \institution{Texas A\&M University}
  \city{College Station, Texas}
  \country{USA}}
\email{ehassani@tamu.edu}

\author{Md Taufique Hussain}
\affiliation{%
  \institution{Wake Forest University}
  \city{Winston-Salem, North Carolina}
  \country{USA}}
\email{hussaint@wfu.edu}

\author{Ariful Azad}
\affiliation{%
  \institution{Texas A\&M University}
  \city{College Station, Texas}
  \country{USA}}
\email{ariful@tamu.edu}


\begin{abstract}
We present scalable distributed-memory algorithms for sparse matrix permutation, extraction, and assignment. 
Our methods follow an Identify–Exchange–Build (IEB) strategy where each process identifies the local nonzeros to be sent, exchanges the required data, and then builds its local submatrix from the received elements. 
This approach reduces communication compared to SpGEMM-based methods in distributed memory. 
By employing synchronization-free multithreaded algorithms, we further accelerate local computations, achieving substantially better performance than existing libraries such as CombBLAS and PETSc.

We design efficient software for these operations and evaluate their performance on two university clusters and the Perlmutter supercomputer. 
Our experiments span a variety of application scenarios, including matrix permutation for load balancing, matrix reordering, subgraph extraction, and streaming graph applications. In all cases, we compare our algorithms against CombBLAS, the most comprehensive distributed library for these operations, and, in some scenarios, against PETSc. 
Overall, this work provides a comprehensive study of algorithms, software implementations, experimental evaluations, and applications for sparse matrix permutation, extraction, and assignment.
\end{abstract}





\maketitle

\section{Introduction}
Sparse matrices are foundational to many scientific computing and machine learning problems in science and engineering. 
Among the most fundamental operations involving sparse matrices are permutation, submatrix extraction, and submatrix assignment. For instance, random permutations can help balance load in distributed environments~\cite{azad2021combinatorial}, while user-defined permutations may impose desirable structures, such as concentrating nonzeros along the diagonal in sparse linear solvers~\cite{duff2001algorithms}. 
In dynamic and streaming graph applications, batches of edges can be added or removed using matrix extract and assign operations, respectively. 
Due to the broad applicability of these operations, they are supported in nearly all major sparse matrix libraries, including PETSc~\cite{Petsc2014}, CombBLAS~\cite{azad2021combinatorial}, Trilinos~\cite{heroux2005overview}, GraphBLAS~\cite{davis2019algorithm}, and SuperLU~\cite{li2003superlu_dist}, among others.

Algorithms for sparse matrix permutation, extraction, and assignment generally follow one of two approaches:
(1) those that leverage sparse matrix-matrix multiplication (SpGEMM), and
(2) those that directly access and manipulate the relevant nonzeros in the matrix without constructing intermediate matrices.
In the first approach, matrix permutation can be defined as  $\mat{A^\prime} = \mat{P}\mat{A}\mat{Q^T}$, where $\mat{P}$ and $\mat{Q}$ are Boolean row and column permutation matrices and $\mat{A}$ is the original matrix.
Similarly, matrix extract and assign operations can be expressed as combinations of two SpGEMM operations~\cite{bulucc2012parallel}.
Buluc and Gilbert proposed efficient SpGEMM-based indexing algorithms, which were implemented in the CombBLAS library~\cite{bulucc2012parallel, azad2021combinatorial}.
The second approach performs these operations by directly communicating the necessary data using collective or point-to-point communication. 
We call this approach {\em Identify-Exchange-Build} (IEB), where each process identifies the local nonzero elements that need to be sent to other processes, performs the required data exchange, and then builds the local submatrix using the received elements.
Several widely used libraries, such as PETSc~\cite{Petsc2014} and SuperLU\_Dist~\cite{li2003superlu_dist}, implement matrix permutation using this communication-based approach. 

While SpGEMM-based algorithms are asymptotically efficient~\cite{bulucc2012parallel}, our experiments show that their practical performance is often inferior to that of IEB approaches. The suboptimal performance of SpGEMM-based permutation, assign, and extract operations can be attributed to several factors.
These methods typically require two separate SpGEMM operations, which can increase both communication overhead and memory usage.
Furthermore, general-purpose SpGEMM implementations are not optimized for permutation and indexing tasks. 
For instance, in matrix permutation, the permutation matrix is extremely sparse,  containing only one nonzero per row or column. Distributed SpGEMM algorithms that are effective for multiplying matrices with comparable sparsity patterns are not well-suited for such highly imbalanced sparsity.
As an example, CombBLAS implements matrix permutation using Sparse SUMMA~\cite{bulucc2012parallel}, which keeps the output matrix stationary while broadcasting submatrices of the input matrices. 
However, when one of the matrices (e.g., the permutation matrix) is extremely sparse, much of the communicated data may be unused, resulting in wasted communication. Consequently, even Sparse SUMMA, one of the best-performing distributed SpGEMM algorithms, performs poorly for permutation, and by extension, for extract and assign operations.

By contrast, IEB-based models can be carefully tailored to exploit the structural characteristics of permutation, extraction, and assignment. 
However, they also present inherent challenges, such as the need for substantial local computation to identify the positions of new elements and to rebuild and sort the local submatrix after data exchange. 
Addressing these challenges requires efficient parallelization strategies at both the process and thread levels, along with optimized communication routines.

This paper presents scalable algorithms for sparse matrix permutation, extraction, and assignment on distributed-memory systems.
Our algorithms are designed under the IEB model and exploit shared computational and communication patterns across these operations.
For communication, our algorithms achieve theoretical lower bounds for matrix permutation, extraction, and assignment operations. 
For local computation, we develop multithreaded algorithms that avoid synchronization among threads to achieve linear thread-level scalability within nodes.
As a result, our algorithms demonstrate strong scalability, running efficiently on up to 4,096 processes (65,536 cores across 512 nodes) of the Perlmutter supercomputer.

Our algorithms enable customized fusion of permutation, extraction, and assignment operations, which are frequently required in applications involving dynamic matrices and graphs. 
We demonstrate this flexibility through an incremental protein clustering application on a large-scale metagenomic dataset. Achieving such fused operations is significantly more difficult with existing approaches.

Overall, this paper makes the following contributions:
\begin{enumerate}
    \item {\bf Distributed-memory algorithms} for sparse matrix permutation, extraction, and assignment based on the Identify-Exchange-Build (IEB) model. These operations share similar computation and communication patterns, enabling a unified and efficient design.
    \item {\bf Scalability:} Our algorithms demonstrate linear thread-level scalability within nodes and excellent parallel efficiency across nodes, scaling up to 4,096 processes (65,536 cores across 512 nodes) on the Perlmutter supercomputer.
    \item {\bf Application-driven evaluation:} We evaluated our algorithms across various application scenarios, including load-balanced permutation, bandwidth-reducing reordering, cluster extraction from graphs, and streaming graph updates. We also integrated multiple sparse matrix operations into an incremental protein clustering application, showcasing both the flexibility and performance of our approach.
\end{enumerate}

The source code and supplementary materials are available in our public GitHub repository\footnote{\url{https://github.com/HipGraph/HipPerm}}. We intend to integrate our code into CombBLAS after further tests.

\begin{figure}[!t]
\centering
\includegraphics[width=\textwidth]{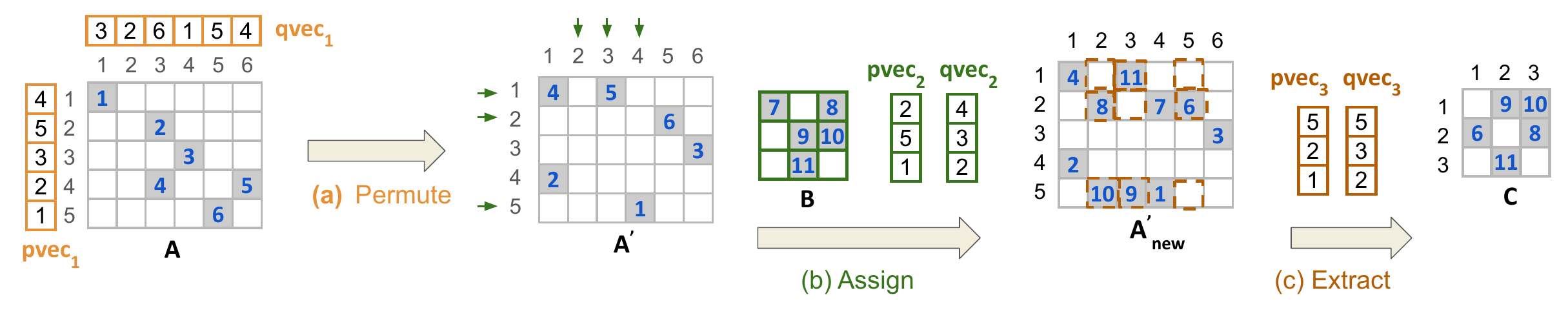}
\caption{Toy examples of matrix operations.
(a) Permutation: Matrix $\mat{A}$ is permuted by the row permutation vector $\vect{pvec_1}$ and column permutation vector $\vect{qvec_1}$ to form matrix $\mat{A^\prime}$. The permutation shuffle the elements of matrix $\mat{A}$ such that $\mat{A} [\vect{pvec_1}[i], \vect{qvec_1}[j]] = \mat{A}^\prime [i,j]$.
(b) Assign: Matrix $\mat{B}$ is assigned into matrix $\mat{A^\prime}$ at the row and column positions specified by the row index vector $\vect{pvec_2}$ and the column index vector $\vect{qvec_2}$. 
The rows and columns of $\mat{A^\prime}$ to be modified by this assignment are marked with arrowheads.
$\mat{A^\prime_{new}}$ is the result of this assignment where $\mat{A^\prime_{new}}[\vect{pvec_2}[i], \vect{qvec_2}[j]] = \mat{B}[i,j]$ for all nonzero indices $i$ and $j$ of $\mat{B}$.
For example, $\mat{B}[1,1]=7$ is assigned to $\mat{A^\prime_{new}}[\vect{pvec_2}[1],\vect{qvec_2}[1]]=\mat{A^\prime_{new}}[2,4]$. 
(c) Extract: A submatrix is extracted from selected locations of matrix $\mat{A^\prime_{new}}$ using the row and column indices specified by the vectors $\vect{pvec_3}$ and $\vect{qvec_3}$, respectively.
Each element $\mat{C}[i,j]$ in the extracted submatrix is defined as
$\mat{A^\prime_{new}}[\vect{pvec_3}[i], \vect{qvec_3}[j]]$ for all $(i,j) \in \vect{pvec_3}\times \vect{qvec_3}$.
Entries extracted from $\mat{A^\prime_{new}}$ are marked with dashed squares in $\mat{A^\prime_{new}}$. 
}
\label{fig-general}
\end{figure}

\section{Problem Formulation and Related Work}
\subsection{Permuting a sparse matrix}
\label{sec:perm_def}
Given a sparse matrix $\mat{A} \in \mathbb{R}^{m\times n}$, we consider a row permutation vector $\vect{pvec}$ representing a permutation of $(1, 2, \ldots, m)$, and a column permutation vector $\vect{qvec}$ representing a permutation of $(1, 2, \ldots, n)$.
Then, the permuted matrix $\mat{A^\prime} \in \mathbb{R}^{m\times n}$ shuffles nonzero entries in $\mat{A}$ such that $\mat{A^\prime}[i,j]=\mat{A}[\vect{pvec}[i], \vect{qvec}[j]]$. 
Matrix permutation can also be expressed using matrix multiplication. Let $\mat{P}$ be a $m\times m$ sparse Boolean matrix such that $\mat{P}[i,j] = 1$ if $j = \vect{pvec}[i]$, and $\mat{P}[i,j] = 0$ otherwise.
Similarly, let $\mat{Q}$ be a $n\times n$ sparse Boolean matrix such that $\mat{Q}[i,j] = 1$ if $j = \vect{qvec}[i]$, and $\mat{Q}[i,j] = 0$ otherwise.

Then, matrix permutation can be defined as $\mat{A^\prime} = \mat{P}\mat{A}\mat{Q^T}$. 
For square matrices, the same permutation can be applied to both rows and columns in the form $\mat{P}\mat{A}\mat{P}\transpose$. 
In the matrix formulation, matrix permutations can be implemented using sparse matrix-matrix multiplication (SpGEMM).
Figure \ref{fig-general}(a) illustrates an example of permuting $\mat{A}$ by two randomly generated permutation vectors.

{\bf Variants of matrix permutations.}
Different applications permute matrices for various purposes, resulting in multiple variants of this operation.
In parallel applications, a randomly generated permutation vector typically redistributes nonzeros throughout the matrix, helping to balance computational and memory loads evenly across processors.
On the other hand, a user-defined permutation can impose a desired structure, such as concentrating nonzeros along the diagonal~\cite{balaji2023community,trotter2023bringing}.
For example, we can permute a matrix to reduce its bandwidth in iterative solvers or ensure a zero-free diagonal during the pivoting step of direct solvers. 
Various algorithms can be used to obtain such orderings, including Reverse Cuthill-McKee (RCM)~\cite{cuthill1969reducing, azad2017reverse}, minimum-degree ordering~\cite{george1989evolution}, and bipartite matching~\cite{duff2001algorithms, azad2020AWPM}.
When $\mat{A}$ represents the adjacency matrix of a graph, we use a symmetric permutation, which means that $\vect{pvec}=\vect{qvec}$. 
In this setting, the permutation reorders the nodes of the graph.

\subsection{Extracting a submatrix from a matrix}
\label{sec:extract_def}
The extract operation retrieves a submatrix from a larger matrix $\mat{A}\in{\mathbb{R}^{m\times n}}$ based on specified subsets of row and column indices.
Let $\vect{pvec}$ be a set of $m^\prime$ row indices selected from the set of all row indices $\{1, 2, \ldots, m\}$ and $\vect{qvec}$ be a set of $n^\prime$ row indices selected from the set of all column indices $\{1, 2, \ldots, n\}$.
The extracted submatrix $\mat{B}\in{\mathbb{R}^{m^\prime\times n^\prime}}$ includes nonzeros from $\mat{A}$ selected by Cartesian products of $\vect{pvec}$ and $\vect{qvec}$: $\mat{B}[i,j] = \mat{A}[\vect{pvec}[i], \vect{qvec}[j]]$ for all $(i,j) \in \vect{pvec}\times \vect{qvec}$.
A matrix multiplication-based method can also perform submatrix extraction, formulated as $\mat{B} = \mat{P}\mat{A}\mat{Q}^T$. 
Here $\mat{P}\in \mathbb{R}^{m^\prime\times m}$ is a sparse Boolean matrix where $\mat{P}[i,j]=1$ if $j=\vect{pvec}[i]$ and $\mat{P}[i,j]=0$ otherwise. Similarly, $\mat{Q}\in \mathbb{R}^{n^\prime\times n}$ is a sparse Boolean matrix where $\mat{Q}[i,j]=1$ if $j=\vect{qvec}[i]$ and $\mat{Q}[i,j]=0$ otherwise~\cite{bulucc2012parallel}.
The submatrix extraction is achieved through a two matrix multiplication: first, $\mat{P}\mat{A}$ selects the target rows, and then $(\mat{P}\mat{A})\mat{Q}^T$ extracts the desired columns from that intermediate result of the first multiplication.
Figure \ref{fig-general}(c) demonstrates an example of extracting a submatrix corresponding to the selecting vectors $\vect{pvec}$ and $\vect{qvec}$.

{\bf Variants of matrix extract.} 
The row and column indices in $\vect{pvec}$ and $\vect{qvec}$ are not required to be unique. For instance, it is possible to extract the same row multiple times if it appears more than once in $\vect{pvec}$. 
However, in some applications, the indices must be unique. For example, if $\mat{A}$ represents the adjacency matrix of a graph, then the extracted matrix $\mat{B}$ corresponds to the adjacency matrix of a subgraph defined by the node indices in $\vect{pvec}$ and $\vect{qvec}$. 
In this case, both $\vect{pvec}$ and $\vect{qvec}$ contain unique values.
Additionally, the indices in $\vect{pvec}$ and $\vect{qvec}$ may or may not be ordered. When the indices are sorted in ascending order, the extracted submatrix preserves the original row and column order. When the indices are unordered, the resulting submatrix permutes the rows and columns according to the order specified in $\vect{pvec}$ and $\vect{qvec}$. This behavior allows the extract operation to serve as a general mechanism for matrix permutation. In particular, if $\vect{pvec}$ and $\vect{qvec}$ represent full permutations of the row and column index sets, the extract operation effectively performs a matrix permutation. Thus, matrix permutation can be viewed as a special case of submatrix extraction.

\subsection{Assigning a matrix to a larger matrix}
\label{sec:assign_def}
The assign operation inserts a sparse matrix $\mat{B} \in \mathbb{R}^{m^\prime\times n^\prime}$ into a larger sparse matrix $\mat{A} \in \mathbb{R}^{m\times n}$ where $m^\prime \leq m$ and $n^\prime \leq n$.
The location where $\mat{B}$ is assigned within $\mat{A}$ is specified by two vectors, $\vect{pvec}$ and $\vect{qvec}$, which contain distinct row and column indices of $\mat{A}$, respectively. 
The elements of $\mat{B}$ are placed at the positions in $\mat{A}$ indicated by these indices, such that $\mat{A}[\vect{pvec}[i], \vect{qvec}[j]] = \mat{B}[i, j]$ for all nonzero indices $i$ and $j$.
Matrix assignment can be done through matrix multiplication and a user-defined addition operation, indicated as $\mat{A}=\mat{A}+\mat{P}\mat{B}\mat{Q}^T$ where $\mat{P}\in \mathbb{R}^{m\times m^\prime}$ is sparse Boolean matrix where $\mat{P}[i,j]=1$ if $i=\vect{pvec}[j]$ and $\mat{P}[i,j]=0$ otherwise.
Likewise, $\mat{Q}\in \mathbb{R}^{n\times n^\prime}$ is a sparse Boolean matrix where $\mat{Q}[i,j]=1$ if $i=\vect{qvec}[j]$ and $\mat{Q}[i,j]=0$ otherwise~\cite{bulucc2012parallel}.
The choice of the addition operator is application-specific. It includes arithmetic addition, the select-second operator, and a logical OR (union) operator. 
Figure \ref{fig-general}(b) demonstrates an example of operation on sample matrices.


{\bf Variants of matrix assignment.} 
In the most common variant of the assign operation, any existing value at position $\mat{A}[\vect{pvec}[i], \vect{qvec}[j]]$ is overwritten by the corresponding value from $\mat{B}$. 
However, in another variant, the values are added instead, following the rule $\mat{A}[\vect{pvec}[i], \vect{qvec}[j]] = \mat{A}[\vect{pvec}[i], \vect{qvec}[j]] + \mat{B}[i, j]$. This additive form is useful for updating sparse matrices, as seen in multifrontal methods~\cite{liu1992multifrontal}. 
In dynamic and streaming graphs, the subgraph assign operation is equivalent to adding a batch of new edges to a graph. 
Let $\mat{A}$ and $\mat{B}$ be the adjacency matrices of graphs $G_\mat{A}$ and $G_\mat{B}$, respectively. 
Assigning $\mat{B}$ to $\mat{A}$ using row and column index vectors $\vect{pvec}$ and $\vect{qvec}$ corresponds to adding each edge $(i, j)$ in $G_\mat{B}$ as an edge $(\vect{pvec}[i], \vect{qvec}[j])$ in $G_\mat{A}$.

\section{Related Work}

Classic applications of sparse matrix permutation are found in numerical solvers~\cite{davis2016survey,davis2010algorithm}.
When solving large systems of equations, one often encounters large sparse matrices that require permutation to enhance numerical stability and computational efficiency. 
For instance, Reverse Cuthill–McKee (RCM) algorithm attempts to pull non-zero elements toward the diagonal, thus producing bandwidth-minimized sparse matrix~\cite{cuthill1969reducing}.
Or the Approximate Minimum Degree (AMD) algorithm attempts to permute the sparse matrix for fill-in reduction~\cite{amestoy1996approximate}.
Consequently, most numerical solvers and numerical linear algebra libraries include support for sparse matrix permutation.
In addition to numerical applications, sparse matrix permutation plays a significant role in graph algorithms. 
Here, the arrangement of rows and columns is altered to ensure an even distribution of work during graph processing. 
Beyond permutation, the extraction or assignment of submatrices is also utilized in the context of streaming batch updates to a graph~\cite{hassani2024batch}. 
As a result, these operations are also supported by various graph analysis frameworks.
We review related work from both parallelization and application perspectives, covering shared-memory algorithms, distributed-memory algorithms, and graph update frameworks.

{\bf Distributed memory frameworks.} 
CombBLAS~\cite{azad2021combinatorial} is an MPI-based distributed memory graph library that leverages 2D/3D partitioning of sparse matrices to minimize communication overhead and optimize performance~\cite{bulucc2012parallel}. 
It supports different functions like sparse matrix permutation and reordering, extracting a submatrix, and matrix assignment using an SpGEMM-based implementation.
Trilinos~\cite{heroux2005overview, trilinos-website} is another well-known software framework designed for scalable scientific computing, featuring a variety of linear algebraic packages. 
Packages like Tpetra and Epetra~\cite{tpetra-website, epetra-website} with Kokkos can be adapted for tasks such as submatrix extraction, matrix permutation, and assignment. 
Both Tpetra and Epetra packages, follow IEB-like approach where Export/Import objects perform communication between the processes in a particular distribution pattern call Maps, and subclasses define how to pack elements into a buffer for sending with MPI, and how to unpack received elements. 
Additionally, for assigning functions, {\tt SumIntoGlobalValues} or {\tt ReplaceGlobalValues} is used to update entries in local matrix classes.
While Trilinos is primarily a 1D row block distributed framework, it can also be adapted for 2D distribution problems~\cite{mayr2025trilinos}.

Another distributed memory library PETSc~\cite{petscsf2022} is designed for scientific computing focused on numerical solutions. 
PETSc’s data structures and algorithms are capable of different sparse matrix computations including permutation, reordering for special structures, submatrix extraction and assignment.
This framework uses 1D row block distribution MPI with support for inter process parallelization with MPI and local parallelization through OpenMP.
However, not all functions in PETSc is OpenMP parallel.
PETSc permute function {\tt MatPermute} reorders a sparse matrix by creating a new matrix that incorporates the new ordering by performing necessary inter process communication. 
Extracting a submatrix with {\tt MatCreateSubMatrix} function also involves communication between processes to gather non-local data required for creating a submatrix. 
To assign a matrix in PETSc, {\tt MatSetValues} is used to assign new elements to any location of the global matrix. 
For elements that belongs to a different process, PETSc buffers the values locally until {\tt MatAssemblyBegin} is called, followed by {\tt MatAssemblyEnd}.  
These functions involve collection communication to ensure all buffered data is exchanged. 

SuperLU~\cite{demmel1999superlu,li2005overview} is a general purpose library for the direct solution of sparse linear systems.  
SuperLU mainly provides functions for LU factorization of sparse matrices along with related routines for matrix reordering.
This library provides separate implementations optimized for distributed memory systems, SuperLU\_DIST~\cite{li2003superlu_dist}, shared-memory multicore processors, SuperLU\_MT~\cite{demmel1999asynchronous}, GPU-accelerated platforms through the SuperLU\_GPU extension~\cite{li2023newly}. 
SuperLU's design avoids explicit matrix permutation due to its complexity. 
However, in {\tt Preprocessing} step of its distributed memory algorithms on matrix $\mat{A}$, permutation vectors ${\vect{P_r}}$ and ${\vect{P_c}}$ computed such that ${\vect{P_r}}A{\vect{P_c}}$ represent a more numerically stable and sparser matrix suitable for LU factorization algorithms.
Then in further steps of algorithms, ${\vect{P_r}}$ and ${\vect{P_c}}$ guide the access pattern for better locality~\cite{li2023newly,li1998making}.

Cyclops Tensor Framework (CTF)~\cite{solomonik2014massively} is a distributed memory numerical library designed for parallel computation on tensors.
It supports a variety of data distribution models, including 1D, 2D, and 3D, and leverages a hybrid MPI+OpenMP parallelism model.
Indexing operations, such as permutation, extraction, and assignment, are implemented in CTF using a method conceptually similar to the IEB model.
Each process first identifies the locations of elements to be manipulated, then redistributes the data, and finally performs the local computation to update values.
However, CTF uses the index mapping vectors for subsequent computations instead of explicitly relocating elements or creating a new submatrix, thus delaying communication until it is strictly necessary~\cite{solomonik2013cyclops}.

{\bf Shared-memory frameworks.} 
GraphBLAS~\cite{davis2019algorithm,davis2023algorithm} is a shared memory linear algebraic graph framework that provides sparse matrix operations to support parallel graph algorithms. 
This library supports core indexing operations such as permutation, subgraph extraction, and graph assignment. 
Permutation and submatrix extraction are achieved using {\tt GrB\_extract} primitive which creates a new matrix.
This operation is done in multiple phases to ensure work is evenly divided into parallel tasks, leading to better parallelism. 
In GraphBLAS a sparse matrix can be updated with another matrix using the assignment operation {\tt GrB\_assign}.
{\tt GrB\_assign} operation postpones changes to the matrix data structure, buffering new or modified entries in an intermediate format until a subsequent phase or explicit call.
Parallel implementations of these operations are highly optimized and outperform MATLAB when multiple OpenMP threads are used~\cite{aznaveh2020parallel}. 
GraphBLAST~\cite{yang2022graphblast} is the GPU implementation of GraphBLAS, which supports all three indexing operations mentioned.

Galois~\cite{nguyen2013lightweight} is another shared memory graph analytics library designed for parallel implementations of graph operations, supporting irregular graph processing tasks.
Permuting a graph can be achieved using {\tt permute} function in Galois.
Galois has a direct approach to remap matrix locations for permutation, and it is OpenMP parallel. 
Galois doesn't have a specific function to extract a submatrix.
Since Galois supports dynamic graph capabilities with its {\tt MorphGraph} data structure, matrix assignment is possible when the graph is represented this way.
The assign operation is not supported for graphs represented with the CSR data structure in Galois.
D-Galois~\cite{dang2018lightweight} is the distributed version, and CUDA-integrated Galois for GPUs extends Galois’s core operations to distributed and GPU environments. 
However, both D-Galois and the GPU version are optimized for specific graph functions and currently lack support for dynamic graph updates or subgraph extraction.

{\bf Batch graph update frameworks.} 
To support batch graph updates, several frameworks employ different data structures and memory models to optimize for graph assignment. 
STINGER~\cite{ediger2012stinger} is a shared memory framework that offers a blocked linked list data structure suitable for streaming graphs and supports batch and single updates on graphs.
Because STINGER employs a blocked adjacency structure with preallocated memory for vertices it is highly optimized for batch updates without the need for rebuilding the graph adjacency matrix.
DISTINGER~\cite{feng2015distinger} extends STINGER’s capabilities to distributed memory systems, using MPI for communication. 
For GPU environments, cuSTINGER~\cite{green2016custinger} and Hornet~\cite{8547541} both leverage GPU parallelism, with cuSTINGER adapting STINGER’s features with different data structures of arrays for the neighbor lists and Hornet introducing a data structure of compressed adjacency list on dynamic graphs. 
Aspen~\cite{dhulipala2019low} is another shared memory graph streaming system with tree based data structure which supports batch update of graphs with high throughput.
SSTGraph~\cite{wheatman2021streaming}, a shared memory parallel framework, leverages a dynamic set data structure for streaming graphs and supports batch graph update through sort merge method.
GraphIn~\cite{sengupta2016graphin} is a distributed framework for incremental graph processing that handles updates and querying of the dynamic graph. 
Its data structure is hybrid with compressed matrix format for static and edgelist for updates.

Although batch updates can achieve higher throughput, many streaming and dynamic graph frameworks must sort batches before updating their graph data structure. 
This sorting impacts the performance of the batch updates~\cite{wheatman2021streaming,dhulipala2019low,pandey2021terrace}.

\section{HipPerm: High-performance Matrix Permutations in Distributed Memory}
\subsection{Data Distribution}
Distributed matrix computations typically partition matrices across processes using 1D (row/column), 2D (block/checkerboard), or 3D (layers of 2D partitions) layouts.
Each of these partitioning techniques comes with distinct communication and memory footprints. 
One-dimensional partitions often scale poorly at large process counts, particularly for irregular and highly imbalanced sparsity where nonzeros are concentrated in only a few rows or columns.
Two-dimensional layouts mitigate these issues by distributing nonzeros more evenly across processes, reducing memory and communication bottlenecks. 
Three-dimensional schemes are used mainly for matrix–matrix multiplication to lower communication at the cost of additional replication and complexity, and are not general-purpose for arbitrary sparse kernels.
Given the general-purpose nature of permutation, extraction, and assignment operations, we adopt a 2D partitioning of matrices and vectors for all algorithms presented in this paper.

\begin{table}[!t]
    \centering
    \caption{Symbols and notations.}
    \label{tab:notation}
    \resizebox{0.95\linewidth}{!}{
    \begin{tabular}{llll}
        \toprule
        \textbf{Notation} & \textbf{Meaning} & \textbf{Notation} & \textbf{Meaning} \\
        \toprule   
        $p$ & Number of processes in distributed setting & $\mathcal{O}(.)$ & Asymptotic complexity\\
        $P_{r,c}$ & The $r$-th row and $c$-th column process in process grid & $\mat{A}[i,j]$ & Element in $i$-th row and $j$-th column of matrix $\mat{A}$  \\
        $nnz(A)$ & The number of nonzeros in sparse matrix $\mat{A}$ & $\mat{A}_{r,c}$ & Submatrix of A stored in process $P_{r,c}$\\
        $|pvec|$ & length of vector $\vect{pvec}$ & $\alpha$ & Latency cost of communication\\
        $\beta$ & Cost per byte in communication & $\texttt{nthreads}$ & Number of OpenMP threads per process\\
        \bottomrule
    \end{tabular}
    }
\end{table}

We use a 2D distribution of the data across $p$ processes in a $\sqrt{p}\times \sqrt{p}$ process grid.  
The matrices are divided into 2D block partitions, and each submatrix block is stored in one process in the grid.  
We denote the process and submatrix in $r$-th row and $c$-th column of the process grid with $P_{r,c}$ and $\mat{A}_{r,c}$, respectively.
For matrix $\mat{A}\in \mathbf{R}^{m\times n}$, each process $P_{r,c}$ stores submatrix $\mat{A}_{r,c}$ with dimension $m/\sqrt{p}\times n/\sqrt{p}$. 
Vectors are also partitioned among all $p$ processes without any replication. 
We denote the subvector of the vector $\vect{vec}$ stored on $P_{r,c}$ with $\vect{vec}_{r,c}$. 
Considering a $\sqrt{p}\times \sqrt{p}$ process grid, the vector $\vect{vec}$ is partitioned in a way that processes in row grid $r$ collectively store subvector $\vect{vec}(r,:)$ with the length of $\lfloor \frac{\lvert \vect{vec}\lvert}{\sqrt{p}} \rfloor$. 
Within the same row, the subvector $\vect{vec}(r,:)$ is further divided between $\sqrt{p}$ process, such that each process holds a subvector of length $\lfloor \frac{\lvert \vect{vec}(r,:) \lvert}{\sqrt{p}} \rfloor$. 

We denote the number of nonzeros in a matrix 
$\mat{A}$ as $nnz(\mat{A})$.
When the average number of nonzeros per column is on the same order as the number of processes, some local submatrices may contain columns with no nonzero entries. Such matrices are referred to as hypersparse matrices.
To efficiently store such matrices, we use the doubly compressed sparse column (DCSC) format, an adaptation of the standard compressed sparse column (CSC) structure that stores only the columns containing at least one nonzero element~\cite{buluc2008representation}.
All algorithms discussed in this paper use the same 2D data distribution and DCSC data structures.
Table~\ref{tab:notation} summarizes common notations used in the paper.

\begin{figure}[!tp]
\centering
\includegraphics[width=\textwidth]{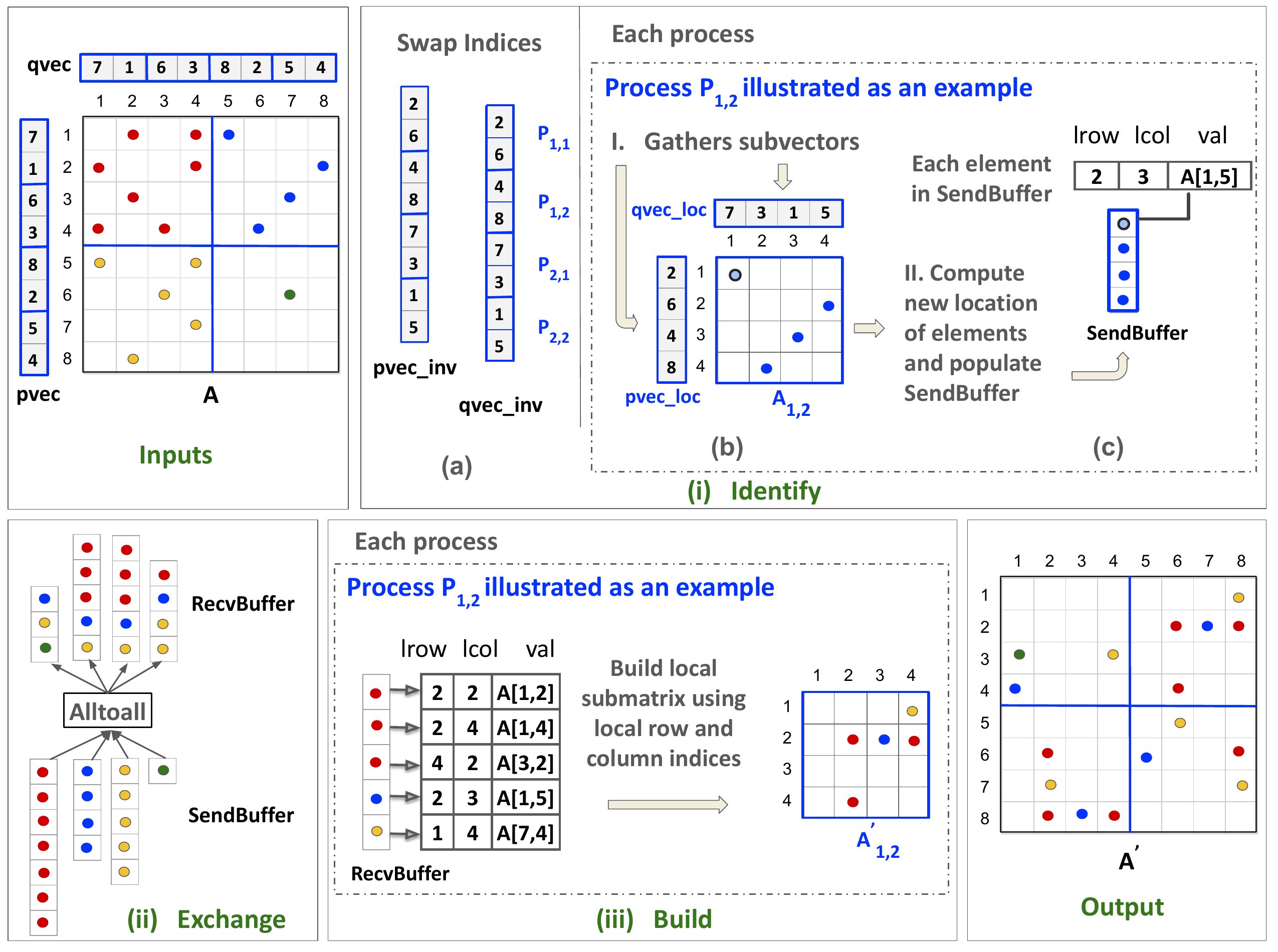}
\caption{An example of the {\em Identify-Exchange-Build} approach applied to a symmetric random permutation of matrix $\mat{A}$.
The top-left panel shows the input matrix along with identical row and column permutation vectors, $\vect{pvec}$ and $\vect{qvec}$.
The inputs are distributed across four processes, with process boundaries indicated by bold blue lines.
(i) In the {\em Identify} step, we first swap indices and values of $\vect{pvec}$ (and similarly for $\vect{qvec}$) such that $ \vect{pvec}[\vect{pvec\_inv}[i]] = i$. 
Next, each process gathers $\vect{pvec\_inv}$ along the process grid row and $\vect{qvec\_inv}$ along the process grid column to obtain all necessary indices. 
In our example, process $P_{1,2}$ gathers $\vect{pvec\_inv}$ from $P_{1,1}$ and $P_{1,2}$ and stores them in $\vect{pvec\_loc}$.
After gathering indices, each process determines the destination of local nonzeros of $\mat{A}$ and groups them by destination process ID into a \texttt{SendBuffer}.
Each nonzero in \texttt{SendBuffer} is stored as a triple $(lrow, lcol, val)$, where $lrow$ and $lcol$ are local indices in the destination submatrix and $val$ is the nonzero value. For example, entry $(1,1)$ in $\mat{A}_{1,2}$ is relocated to global position $(2,7)$, corresponding to local indices $(2,3)$ in $P_{1,2}$.
(ii) In the Exchange step, processes use \texttt{Alltoallv} to exchange \texttt{SendBuffer} data, placing the results in their \texttt{RecvBuffer}. In subfigure (ii), colors in \texttt{RecvBuffer} indicate nonzeros received from different source processes.
(iii) In the Build step, each process reconstructs its local submatrix of $\mat{A}^\prime$ from the \texttt{RecvBuffer} by sorting entries by $(lrow, lcol)$. In  subfigure (iii), $P_{1,2}$ builds $\mat{A}^\prime_{1,2}$ from its received data.
The permuted matrix $\mat{A}^\prime$ is shown in the bottom right subfigure. }
\label{fig:permute}
\end{figure}

\subsection{The HipPerm Algorithm}
As defined in Section~\ref{sec:perm_def}, the permutation of a matrix is defined as  $\mat{A^\prime}[i,j]=\mat{A}[\vect{pvec}[i], \vect{qvec}[j]]$, where the input matrix, permutation vectors and the permuted matrix are distributed on a 2D process grid. 
In a distributed setting, the nonzeros $\mat{A}[\vect{pvec}[i], \vect{qvec}[j]]$ and $\mat{A'}[i,j]$ may reside on different processes. 
Therefore, each process must determine the destination of its local nonzeros, exchange data with other processes accordingly, and reconstruct the permuted matrix after communication.
We refer to this three-step approach as {\em Identify-Exchange-Build}: the first step {\em identifies} the destination process for each local nonzero; the second step performs data {\em exchange} using an 
\texttt{Alltoallv} communication; and the third step {\em builds} the permuted matrix locally on each process.

\begin{algorithm}[!t]
\caption{HipPerm: Permuting a distributed sparse matrix}
\label{algo:hipperm}
\textbf{Input and Output:} Input sparse matrix $\mat{A} \in \mathbb{R}^{m \times n}$, permutation vectors $\vect{pvec}$ and $\vect{qvec}$. Output matrix $\mat{A^\prime} \in \mathbb{R}^{m \times n}$. All matrices and vectors are distributed on a $\sqrt{p}\times \sqrt{p}$ process grid.
\begin{algorithmic}[1]
\Procedure{HipPerm}{$\mat{A}, \vect{pvec}, \vect{qvec}$}
\State $\vect{pvec\_inv}$, $\vect{qvec\_inv}$ $\gets$ SwapIndexValue($\vect{pvec}$, $\vect{qvec}$) \Comment{ Collective communication across processes}
\State $\vect{pvec\_loc}, \vect{qvec\_loc} \gets$ Gather corresponding subvectors of $\vect{pvec}$ and $\vect{qvec}$  

 \For{Every submatrix $\mat{A}_{r,c}$ in process $P_{r,c}$ in \textbf{parallel}} \Comment{MPI parallel}
 \State \SendBuffer $\gets \Call{PrepareSendBuffer}{\mat{A}_{r,c},\vect{pvec\_loc}, \vect{qvec\_loc}}$
 \EndFor
\State $\texttt{RecvBuffer} \gets \Call{MPI\_Alltoallv}{\texttt{SendBuffer}}$ 
\For{Every process $P_{r,c}$ in \textbf{parallel}}
    \State $\mat{A^\prime}_{r,c} \gets \Call{BuildLocalMatrix}{\texttt{RecvBuffer}}$
\EndFor
\State \Return $\mat{A^\prime}$
\EndProcedure
\end{algorithmic}
\end{algorithm}

Our distributed permutation algorithm, HipPerm, is based on the three phases of the Identify–Exchange–Build framework.
We describe these steps using Figure~\ref{fig:permute} and Algorithm~\ref{algo:hipperm}.
Here, HipPerm takes a sparse matrix $\mat{A}$ and two permutation vectors $\vect{pvec}$ and  $\vect{qvec}$ as inputs. All inputs and outputs are distributed on a 2D process grid. 

\subsubsection{Step 1: Identify Permuted Indices and Values}
\label{Indentify-step}
The first step of HipPerm is to identify the destination process for each local nonzero and populate a buffer for data exchange.
We subdivide this step into three substeps.

{\bf(a) Gather necessary indices.} 
By definition of permutation, the entry $\mat{A}[\vect{pvec}[i], \vect{qvec}[j]]$ must be sent to the process that owns $\mat{A^\prime}[i,j]$. Because $\vect{pvec}[i]$, $\vect{qvec}[j]$, and $\mat{A}[\vect{pvec}[i], \vect{qvec}[j]]$ may be in different processes, accessing this entry from a local submatrix may require fine-grained remote communication.
To avoid such remote indexing of $\mat{A}$, we swap indices and values of $\vect{pvec}$ (and similarly for $\vect{qvec}$) such that $ \vect{pvec}[\vect{pvec\_inv}[i]] = i$. 
In other words, each element $\vect{pvec\_inv}[i]$ holds the index (position) at which the value $i$ appears in 
$\vect{pvec}$.
Substep (a) in Figure~\ref{fig:permute}(i) illustrates this index-value swapping step. 
In a distributed setting, this inversion requires \texttt{Alltoall} communication for both $\vect{pvec}$ and $\vect{qvec}$.
Next, each process gathers $\vect{pvec\_inv}$ along the process grid row and $\vect{qvec\_inv}$ along the process grid column to obtain the necessary indices to access the local submatrix. 
For example, in Substep (b) of Figure~\ref{fig:permute}(i), the process $P_{1,2}$ gathers $\vect{pvec\_inv}$ from $P_{1,1}$ and $P_{1,2}$ and stores them in $\vect{pvec\_loc}$.
This step performs two \texttt{Allgather} operations across $\sqrt{p}$ processes, storing the resulting subvectors in local memory, as shown in Lines 2 and 3 of Algorithm~\ref{algo:hipperm}.

{\bf(b) Access the local submatrix, identify destination processes, and populate a send buffer.} 
In this substep, process $P_{r,c}$ scans its local submatrix $\mat{A}_{r,c}$ and builds a \texttt{SendBuffer}, as shown in Figure~\ref{fig:permute}(i)(c) and Lines 5–10 of Algorithm~\ref{algo:hipperm}. 
Entries in \texttt{SendBuffer} are grouped by the destination process id.
For example, let \texttt{DestP} be the process that holds $\mat{A}[\vect{pvec\_loc}[i], \vect{qvec\_loc}[j]]$ on the 2D process grid. 
Then, $\mat{A}_{r,c}[i,j]$ will be sent to \texttt{DestP} after permutation. Consequently, $\mat{A}_{r,c}[i,j]$ is grouped with all other local entries that are destined to \texttt{DestP}. 
Each entry of \texttt{SendBuffer} is a triple $(lrow, lcol, val)$, where $lrow$ and $lcol$ are local indices in the destination submatrix and $val$ is the nonzero value. This step requires no communication, costs $O(nnz/p)$, and is the most compute-intensive part of HipPerm. 
We implement it in \textsc{PrepareSendBuffer} (Algorithm~\ref{algo:hipperm}, Line 5) and accelerate it with multithreaded parallelization, discussed in the next section.

\subsubsection{Step 2: Exchange Data}
After each process populates its \texttt{SendBuffer} array, an \texttt{Alltoallv} communication step is performed to exchange nonzero elements between processes as shown in Figure~\ref{fig:permute}(ii).
We will analyze the communication complexity in the next section.

\subsubsection{Step 3: Build the Permuted Matrix}
\label{sec:build-step}
After the \texttt{Alltoallv} exchange, each process independently constructs its local submatrix using the nonzero elements stored in the \texttt{RecvBuffer}, as illustrated in Figure~\ref{fig:permute}(iii). To build the local sparse matrix data structure, we first sort the received tuples by column and row indices. For parallel sorting, the \texttt{RecvBuffer} is divided into $4 \times \texttt{nthreads}$ chunks, and each thread processes one chunk in parallel. The sorted chunks are then merged using a parallel multiway merge algorithm~\cite{azad2016exploiting}. Once sorting is complete, the new local submatrix is constructed in the DCSC format.

\begin{algorithm}[!t]
\caption{Preparing Send Data at process $P_{i,j}$}
\label{algo:threadComputation}
\textbf{Input and Output:} Local matrix $\mat{A}_{r,c}$, local permutation vectors $\vect{pvec\_loc }$ and $\vect{qvec\_loc }$. Output: The \SendBuffer ~ array ready for communication.
\begin{algorithmic}[1]
\Procedure{PrepareSendBuffer}{$\mat{A}_{r,c}, \vect{pvec\_loc}, \vect{qvec\_loc}$}
\State $\nnz \gets $ Number of nonzeros in $\mat{A}_{r,c}$
\State Logically partition $\mat{A}_{r,c}$ by columns into $\nthreads$ parts, each containing approximately $\frac{\nnz}{\texttt{nthreads}}$ nonzeros 
\State $\texttt{nnzCounter} \gets \text{zeros}(\texttt{nthreads} \times p)$ \Comment{Initialize a vector with zero values}

\For {$t \gets 1$ to \nthreads ~  \textbf{in parallel}} \Comment{OpenMP thread parallel}
    \For {Each nonzero at $(i,j)$ in the $t$-th partition of $\mat{A}_{r,c}$} 
    \State \texttt{DestP}$ \gets$ Compute the process that owns $\mat{A}[\vect{pvec\_loc}[i], \vect{qvec\_loc}[j]]$ 
    \State $\texttt{nnzCounter}[t,\texttt{DestP}]\gets \texttt{nnzCounter}[t,\texttt{DestP}] + 1$ 
    \EndFor
\EndFor

\For {$t \gets 1$ to \nthreads ~  \textbf{in parallel}} \Comment{OpenMP thread parallel}
    \State $\Tilde{\mat{A}} \gets$ the $t$-th partition of $\mat{A}_{r,c}$
     \For {Each nonzero at $(i,j)$ in $\Tilde{\mat{A}}$} 
    \State \texttt{DestP}$ \gets$ Compute the process that owns $\mat{A}[\vect{pvec\_loc}[i], \vect{qvec\_loc}[j]]$ 
    \State $(lrow, lcol) \gets $ local row and column indices corresponding to $\vect{pvec\_loc}[i]$ and $\vect{qvec\_loc}[i]$ in  \texttt{DestP}
    \State $index = \Call{SendBufferIndex}{\texttt{DestP}, \texttt{nnzCounter}}$ \Comment{Explained in Figure.~\ref{fig-LocalPartitioning}(ii)}
    \State $\SendBuffer[index] \gets (lrow, lcol, \Tilde{\mat{A}}[i, j])$
    \EndFor
\EndFor
\State \Return \SendBuffer
\EndProcedure
\end{algorithmic}
\end{algorithm}

\begin{figure}[!t]
\centering
\includegraphics[width=.95\textwidth]{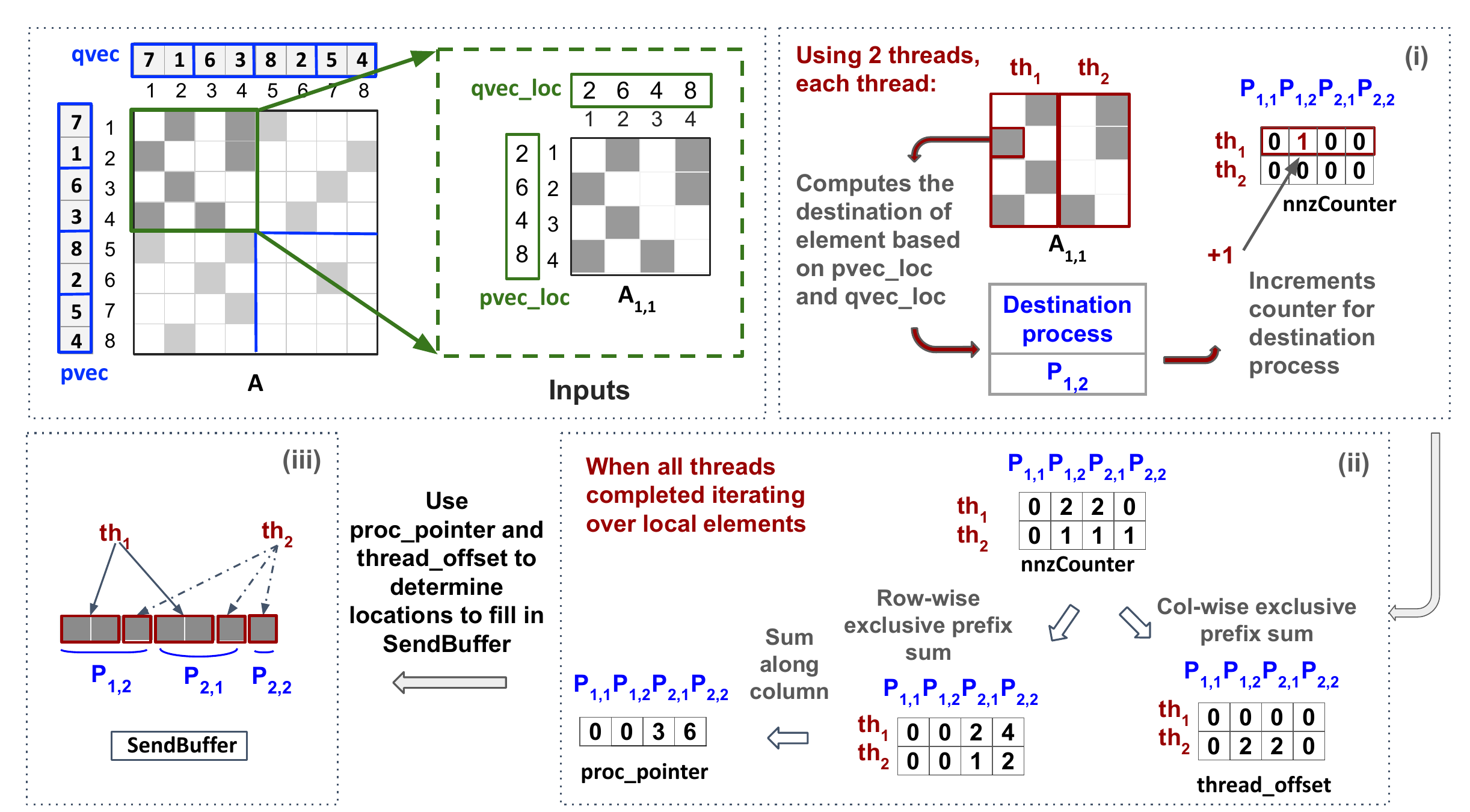}
\caption{
Illustration of the \textsc{PrepareSendBuffer} function within the {\em Identify} step in HipPerm.
This example considers two threads within process $P_{1,1}$, which holds the local submatrix $\mat{A}_{1,1}$ and the inverse permutation vectors $\vect{pvec\_loc}$ and $\vect{qvec\_loc}$.
(i)  $\mat{A}_{1,1}$ is partitioned along the column among two threads (indicated by red lines). Each thread accesses nonzeros in its part of $\mat{A}_{1,1}$ and populates \texttt{nnzCounter}, where \texttt{nnzCounter}[$i,j$] stores the number of nonzeros identified by thread $i$ that are destined for process $j$.
(ii) We apply a row-wise prefix sum on \texttt{nnzCounter}, followed by a column-wise sum to generate the \texttt{proc\_pointer} vector. A column-wise prefix sum creates the \texttt{thread\_offset} array that is used as an offset thread $t$ writes data destined for process $i$.
(iii) In this step, each thread populates the \texttt{SendBuffer} by using \texttt{proc\_pointer} to find the base address for each target process and \texttt{thread\_offset} to determine its specific write location.
}

\label{fig-LocalPartitioning}
\end{figure}

\subsection{Multithreaded Algorithms for Local Computations}
As shown in Algorithm~\ref{algo:hipperm}, the HipPerm algorithm includes two local computation functions: \textsc{PrepareSendBuffer} and \textsc{BuildLocalMatrix}.
Among these, \textsc{PrepareSendBuffer} is by far the most computationally intensive, as it involves accessing all local nonzeros on each process $P_{i,j}$, determining their destination processes, and preparing the buffer for data exchange.
Our approach to parallelizing these local computations within each node is illustrated in Algorithm~\ref{algo:threadComputation} and Figure~\ref{fig-LocalPartitioning}.

The primary objectives of our multithreaded local computation are: (a) to achieve load-balanced execution across threads, and (b) to eliminate thread synchronization entirely, enabling linear scalability within each node.
To ensure balanced workload across the $\nthreads$ threads, we partition the local submatrix by columns into $\nthreads$ parts, each containing approximately $\frac{\nnz}{\nthreads}$ nonzeros.
Each thread then processes one part independently and in parallel.

To eliminate thread synchronization when populating the shared send buffer, we process the local submatrix $\mat{A}_{r,c}$ in two passes.
In the first pass, each thread counts how many nonzero elements need to be sent to each process.
This step fills a $\texttt{nthreads} \times p$ array called \texttt{nnzCounter}, where \texttt{nnzCounter}[$i,j$] stores the number of nonzeros identified by thread $i$ that are destined for process $j$, as shown in line 5-10 in Algorithm~\ref{algo:threadComputation} and Figure~\ref{fig-LocalPartitioning}(i).

We use \texttt{nnzCounter} to compute unique write indices into \texttt{SendBuffer}, allowing all threads to write without conflicts.
First, we apply a row-wise prefix sum on \texttt{nnzCounter}, followed by a column-wise sum to generate the \texttt{proc\_pointer} vector. The data to be sent to the $i$th process is stored starting at $\texttt{SendBuffer}[\texttt{proc\_pointer}[i]]$.
Since multiple threads may simultaneously process data destined for the same process, concurrent writes to \texttt{SendBuffer} can lead to conflicts. To avoid this, each thread computes a specific offset within the range \texttt{SendBuffer}[\texttt{proc\_pointer}[$i$]] to \texttt{SendBuffer}[\texttt{proc\_pointer}[$i+1$]] where it can safely write its data for process $i$.
These per-thread offsets are computed via a column-wise prefix sum on the \texttt{nnzCounter} array, resulting in the \texttt{thread\_offset} array. For example, \texttt{thread\_offset}[$t,i$] stores the offset where thread $t$ writes data destined for process $i$.
The pointer and offset calculations are illustrated in Figure~\ref{fig-LocalPartitioning}(ii) and are encapsulated in the function \textsc{SendBufferIndex} in Algorithm~\ref{algo:threadComputation}.

After the thread-specific indices in \texttt{SendBuffer} are computed, the $t$th thread processes the nonzeros in the $t$th partition of the local submatrix $\mat{A}_{r,c}$. For each nonzero, it determines the destination process and computes the corresponding local row and column indices in the destination process (lines 14–15 of Algorithm~\ref{algo:threadComputation}).
Finally, using the previously computed indices and offsets, the thread stores the data tuples into the appropriate locations in the \texttt{SendBuffer} array (line 17 of Algorithm~\ref{algo:threadComputation}).
Although this approach involves accessing the data twice, it eliminates the need for thread synchronization, enabling strong thread-level scalability within each process, as demonstrated in the results section.

\subsection{HipPerm Complexity Analysis}
\label{sec:perm_analysis}
We analyze HipPerm's complexity separately for \emph{Identify}, \emph{Exchange}, and \emph{Build} stages. For simplicity, we assume a square input matrix $\mat{A} \in \mathbb{R}^{n \times n}$ and permutation vectors of length $n$.
For communication complexity, we use the $\alpha-\beta$ model~\cite{thakur2005optimization},
where $\alpha$ is the latency constant corresponding to the fixed cost of communicating a message regardless of its size, and $\beta$ is the inverse bandwidth corresponding to the cost of transmitting one word of data. Consequently, communicating a message of $m$ words takes $\alpha + \beta m$ time. 

\textbf{\em Identify: } 
The {\em Identify} step performs two collective communications on the permutation vectors. 
Swapping indices and values of vectors requires an \texttt{Alltoall} involving all $p$ processes, each holding $n/p$ vector elements. 
Considering the pairwise-exchange algorithm for \texttt{Alltoall}, the communication complexity is 
$\mathcal{O}(\alpha (p-1) + \beta\frac{n}{p})$.
The next communication uses an \texttt{Allgather} operation among $\sqrt{p}$ processes to gather necessary indices of the inverted permutation vectors. 
Considering the Ring algorithm for \texttt{Allgather}, the communication complexity is 
$\mathcal{O}(\alpha (\sqrt{p}-1) + \beta\frac{n}{\sqrt{p}})$.

Let $\nnz(\mat{A}_{i,j})$ denote the number of nonzeros in process $P_{i,j}$.
Since Algorithm~\ref{algo:threadComputation} accesses local nonzeros twice in a load-balanced manner across $\nthreads$ threads, the per-process complexity is $\mathcal{O}(\nnz(\mat{A}_{i,j})/\nthreads)$.
If the matrix is evenly distributed, the overall cost is $\mathcal{O}(\nnz/(p*\nthreads))$.
However, if nonzeros are unevenly distributed, the cost is dominated by the process with the largest submatrix.

\textbf{\em Exchange: }
The exchange step performs an \texttt{Alltoallv}, where each process sends $3\nnz(\mat{A}_{i,j})$ data.
Assuming a balanced matrix and pairwise-exchange \texttt{Alltoall}, each process sends and receives $3\nnz/p$ data.
The communication complexity is $\mathcal{O}(\alpha(p - 1) + \beta\nnz/p)$.
It is difficult to derive the communication complexity for imbalanced input and output matrix distributions. 
However, as shown in our experiments, such imbalances typically lead to higher communication costs.

\textbf{\em Build: } 
We assume that each process receives approximately $\nnz/p$ nonzeros.
First, we sort the received data with a time complexity of $\mathcal{O}(\frac{\nnz}{p*\nthreads}\log \frac{\nnz}{p*\nthreads})$.
After sorting, a multiway merge of the sorted partitions is performed. We use a hash-based multiway merge that takes $\mathcal{O}(\frac{\nnz}{p*\nthreads})$.

\begin{figure}[!tp]
\centering
\includegraphics[width=\textwidth]{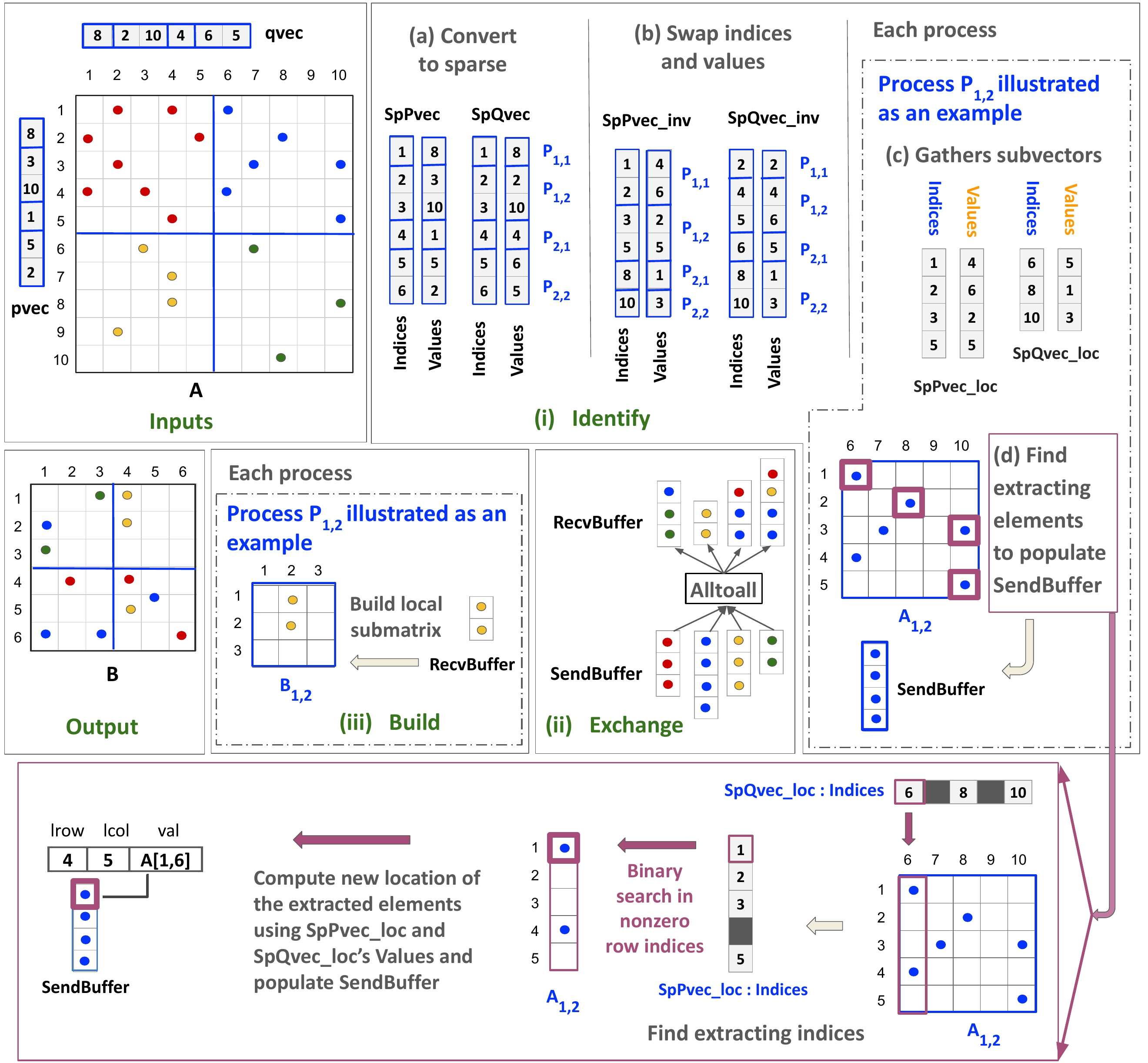}
\caption{An example of {\em Identify-Exchange-Build} method for extracting a submatrix, $\mat{B}\in \mathbb{R}^{|\vect{pvec}|\times |\vect{qvec}|}$, from a larger matrix $\mat{A}\in \mathbb{R}^{n\times n}$. 
The top-left panel presents the input matrix $\mat{A}$ and the selected row and column index vectors, $\vect{pvec}$ and $\vect{qvec}$, distributed on a $2\times2$ process grid. 
Bold blue lines indicate the process boundary.
(i) In the {\em Identify} step, $\vect{pvec}$ and $\vect{qvec}$ are first converted to sparse vectors, and then indices and values are swapped to form $\vect{SpPvec\_inv}$ and $\vect{SpQvec\_inv}$. 
Next, each process gathers $\vect{SpPvec\_inv}$ along the process grid row and $\vect{SpQvec\_inv}$ along the process grid column to obtain all necessary indices.
In step (d) of {\em Identify}, each process searches its local submatrix for relevant nonzero elements based on the $\vect{Indices}$ in $\vect{SpPvec\_loc}$ and $\vect{SpQvec\_loc}$. 
Step (d) of {\em Identify} is illustrated in more detail in the bottom panel. 
For each column index in $\vect{SpQvec\_loc}$, we perform a binary search within that column of the local matrix using the $\vect{Indices}$ of $\vect{SpPvec\_loc}$ to determine whether a nonzero entry exists to be extracted.
For example, in the bottom panel, column 6 of $\mat{A}_{1,2}$ contains two nonzeros, but only the entry at row index 1 is selected since it appears in $\vect{SpPvec\_loc}$.
For this element, the local row and column indices corresponding to the destination process are computed, and the \texttt{SendBuffer} is populated accordingly.
(ii) In the {\em Exchange} step, an \texttt{Alltoallv} communication exchanges the extracted nonzero elements among processes.
(iii) In the {\em Build} step, each process builds its local submatrix with elements in \texttt{RecvBuffer}.
The extracted submatrix $\mat{B}$ is shown in the output.}
\label{fig-Extract}
\end{figure}

\subsection{The HipExtract Algorithm}
As defined in Section~\ref{sec:extract_def}, a submatrix $\mat{B}$ is extracted from a larger matrix $\mat{A}$ as $\mat{B}[i,j] = \mat{A}[\vect{pvec}[i], \vect{qvec}[j]]$, where $\vect{pvec}$ and $\vect{qvec}$ specify the rows and columns extracted from $\mat{A}$.
In a distributed setting, the input matrix $\mat{A}$, the index vectors $\vect{pvec}$ and $\vect{qvec}$, and the output matrix $\mat{B}$ are all distributed across processes. 
Similar to HipPerm, our distributed algorithm HipExtract follows the three steps of the Identify–Exchange–Build approach: (i) each process identifies the relevant local elements using $\vect{pvec}$ and $\vect{qvec}$, (ii) exchanges the selected elements with other processes, and (iii) builds its local portion of $\mat{B}$. 
The key difference from HipPerm lies in the Identify step: HipExtract extracts only the subset of nonzeros specified by $\vect{pvec}$ and $\vect{qvec}$ before determining their new distributed locations. As a result, extracting a small submatrix with HipExtract is less expensive than performing a full permutation with HipPerm.

The HipExtract algorithm is illustrated in Figure~\ref{fig-Extract} and detailed in Algorithm~\ref{algo:hipextract}. 
The three steps of HipExtract are explained as follows:

\subsubsection{Step 1: Identify Indices and Values for Extraction.}
In this step, all processes identify local nonzero elements to be extracted and calculate their destination processes. This step is divided into four substeps as explained in Figure~\ref{fig-Extract}(i).

{\bf Gather necessary indices.} To access $\mat{A}[\vect{pvec}[i], \vect{qvec}[j]]$ for extraction, each process first gathers the relevant indices from $\vect{pvec}$ and $\vect{qvec}$. Unlike in HipPerm, the lengths of $\vect{pvec}$ and $\vect{qvec}$ are much smaller than the matrix dimensions. Therefore, we convert them into sparse vectors $\vect{SpPvec}$ and $\vect{SpQvec}$, where each element is a tuple consisting of the vector index and the corresponding value, as illustrated in substep (a) of Figure~\ref{fig-Extract}(i).
In substep (b), indices and values are swapped to generate $\vect{SpPvec\_inv}$ and $\vect{SpQvec\_inv}$. This swap, which requires an all-to-all communication to exchange vector elements, follows the same procedure described in the HipPerm algorithm section.
Finally, in substep (c) of Figure~\ref{fig-Extract}(i), each process gathers the required subvectors into local sparse vectors, $\vect{SpPvec\_loc}$ and $\vect{SpQvec\_loc}$, to access the indices needed for extraction.

\begin{algorithm}[!t]
\caption{HipExtract: Extracting a submatrix from a larger matrix}
\label{algo:hipextract}
\textbf{Input and Output:} Input sparse matrix $\mat{A} \in \mathbb{R}^{m \times n}$, extracting vectors $\vect{pvec}$ and $\vect{qvec}$. Output matrix $\mat{B} \in \mathbb{R}^{|\vect{pvec}| \times |\vect{qvec}|}$. All matrices and vectors are distributed on a $\sqrt{p}\times \sqrt{p}$ process grid.
\begin{algorithmic}[1]
\Procedure{HipExtract}{$\mat{A}, \vect{pvec}, \vect{qvec}$}
\State $\vect{SpPvec}$, $\vect{SpQvec}$ $\gets$ ConvertToSparse($\vect{pvec}$, $\vect{qvec}$) \Comment{ Collective communication across processes}
\State $\vect{SpPvec\_inv}$, $\vect{SpQvec\_inv}$ $\gets$ SwapIndexValue($\vect{SpPvec}$, $\vect{SpQvec}$) \Comment{ Collective communication across processes}
\State $\vect{SpPvec\_loc}, \vect{SpQvec\_loc} \gets$ Gather corresponding subvectors of $\vect{SpPvec\_inv}$ and $\vect{SpQvec\_inv}$ 

\For{Every submatrix $\mat{A}_{r,c}$ in process $P_{r,c}$ in \textbf{parallel}} \Comment{MPI parallel}
 \State \SendBuffer $\gets \Call{ExtractPrepareSendBuffer}{\mat{A}_{r,c},\vect{SpPvec\_loc}, \vect{SpQvec\_loc}}$ \Comment{Algorithm~\ref{algo:threadComputation-extract} in the Appendix}
\EndFor
\State $\texttt{RecvBuffer} \gets \Call{MPI\_Alltoall}{\texttt{SendBuffer}}$ 
\For{Every process $P_{r,c}$ in \textbf{parallel}}
    \State $\mat{B}_{r,c} \gets \Call{BuildLocalMatrix}{\texttt{RecvBuffer}}$
\EndFor
\State \Return $\mat{B}$
\EndProcedure
\end{algorithmic}
\end{algorithm}

{\bf Extract from local submatrix, identify destination processes, and populate a send buffer.} 
In this substep, each process $P_{r,c}$ identifies the nonzero elements in its local submatrix $\mat{A}_{r,c}$ that need to be extracted to form $\mat{B}$. 
These elements are determined using the row and column indices stored in the $\vect{Indices}$ fields of $\vect{SpPvec\_loc}$ and $\vect{SpQvec\_loc}$, respectively.

The extract algorithm must verify whether the row and column indices in $\vect{SpPvec\_loc}$ and $\vect{SpQvec\_loc}$ exist within the local submatrix $\mat{A}{r,c}$. 
Accordingly, for each column index in $\vect{SpQvec\_loc}$, we perform a binary search within that column of $\mat{A}{r,c}$ using the $\vect{Indices}$ of $\vect{SpPvec\_loc}$ to identify any nonzero entries to be extracted. 
These binary searches are independent and executed in parallel by multiple threads. 
This procedure is illustrated in substep (d) of Figure~\ref{fig-Extract}(i). 
Finally, as shown in the enlarged view of substep (d), each process computes the destination processes for the extracted nonzero element and populates \texttt{SendBuffer}, following the same method as described in HipPerm.
The multithreaded algorithm for this substep is outlined in Algorithm~\ref{algo:threadComputation-extract} in the Appendix.

\subsubsection{Step 2: Exchange Data}
As shown in Figure~\ref{fig-Extract}(ii), an \texttt{Alltoallv} communication step is performed to exchange nonzero elements among processes.

\subsubsection{Step 3: Build the Extracted Submatrix}
After the exchange step, each process forms its local submatrix using the nonzero elements stored in the \texttt{RecvBuffer} independently, as illustrated in Figure~\ref{fig-Extract}(iii). 
The process of building the local submatrix is the same as in HipPerm, as described in Section~\ref{sec:build-step}.

\subsection{HipExtract Complexity Analysis}
We analyze the complexity of HipExtract for each stage of the algorithm separately. For ease of computation, we assume both input and output matrices to be square, such as $\mat{A}\in \mathbb{R}^{n\times n}$ and $\mat{B}\in \mathbb{R}^{m\times m}$. 
Consequently, $|\vect{pvec}|=|\vect{qvec}|=m$.

\textbf{\em Identify: } 
Similar to HipPerm, HipExtract performs an \texttt{Alltoall} to exchange indices and values in $\vect{SpPvec}$ and $\vect{SpQvec}$, producing $\vect{SpPvec\_inv}$ and $\vect{SpQvec\_inv}$. 
Using the pairwise-exchange algorithm, this incurs a communication cost of $\mathcal{O}(\alpha (p-1) + \beta m/p)$. An \texttt{Allgather} then assembles subvectors into $\vect{SpPvec\_loc}$ and $\vect{SpQvec\_loc}$ across $\sqrt{p}$ processes, with the complexity of $\mathcal{O}(\alpha (p-1) + \beta\frac{m}{\sqrt{p}})$.

To extract elements in a local submatrix, each process executes Algorithm~\ref {algo:threadComputation-extract}.
For simplicity, we assume that $\mat{A}$ is distributed evenly among processes such that each local submatrix has $\frac{nnz(\mat{A})}{p}$ nonzero elements.
The maximum number of nonzero columns on each local submatrix is $\frac{n}{\sqrt{p}}$ and each column has $\frac{nnz(\mat{A}) * \sqrt{p}}{n}$ nonzero elements on average.

Suppose we are extracting random row and column indices, the length of the extracting vectors on each process is $|\vect{SpPvec\_loc}|=|\vect{SpQvec\_loc}|=\frac{m}{\sqrt{p}}$.
Based on Algorithm~\ref {algo:threadComputation-extract}, for each index in $\vect{SpQvec\_loc}$, we perform $\frac{m}{\sqrt{p}}$ binary searches in the selected column of the local submatrix. 
As indices in $\vect{SpQvec\_loc}$ are equally dived among \texttt{nthreads} threads, the computational complexity of this step is 
$\mathcal{O}(\frac{m}{\texttt{nthreads}*\sqrt{p}} * \frac{m}{\sqrt{p}} * \log \frac{nnz * \sqrt{p}}{n})$. 

\textbf{\em Exchange: } 
In step 2 of HipExtract, where an \texttt{Alltoall} communication is performed, the communication cost has complexity of $\mathcal{O}(\alpha(p-1)+\beta \frac{nnz(\mat{B})}{p})$. 

\textbf{\em Build: } 
Finally, in {\em Build} step, each process builds the local submatrix with received $\frac{nnz(\mat{B})}{p}$ elements.
Identical to the HipPerm complexity analysis in {\em Build} step of ~\ref{sec:perm_analysis}, sorting \texttt{nthreads} chunks of elements and then applying multiway merge on the sorted chunk takes $\mathcal{O}(\frac{nnz(\mat{B})}{p*\texttt{nthreads}})$.

\subsection{The HipAssign Algorithm}
\label{sec:assign}
As explained in Section~\ref{sec:assign_def}, assigning $\mat{B}\in \mathbb{R}^{m^\prime \times n^\prime}$ to $\mat{A}\in \mathbb{R}^{m\times n}$ means replacing the entries of $\mat{A}$ at rows $\vect{pvec}$ and columns $\vect{qvec}$ with the corresponding elements of $\mat{B}$: $\mat{A}[\vect{pvec}[i], \vect{qvec}[j]] = \mat{B}[i, j]$.
Mathematically, an assignment can be viewed as adding two matrices. 
However, $\mat{A}$ and $\mat{B}$ cannot be added directly due to their different sizes. 
We address this by permuting $\mat{B}$ with $\vect{pvec}$ and $\vect{qvec}$ and expanding it to the dimensions of $\mat{A}$. 
This produces $\mat{B}^\prime \in \mathbb{R}^{m\times n}$, where $\mat{B}^\prime[\vect{pvec}[i], \vect{qvec}[j]] = \mat{B}[i,j]$~\cite{hassani2024batch}. The expanded matrix $\mat{B}^\prime$ is then added to $\mat{A}$ to complete the assignment: $\mat{A} = \mat{A} + \mat{B}^\prime$.
Similar to HipPerm and HipExtract, the distributed assignment algorithm, HipAssign, follows the three steps of the Identify–Exchange–Build approach to compute $\mat{B}^\prime$.
HipAssign then performs one additional step to add two matrices.
Different steps of HipAssign are described in Figure~\ref {fig-Assign} and Algorithm~\ref {algo:hipassign}.
These steps are explained below. 

\subsubsection{Step 1: Identify}
\begin{figure}[!tp]
\centering
\includegraphics[width=\textwidth]{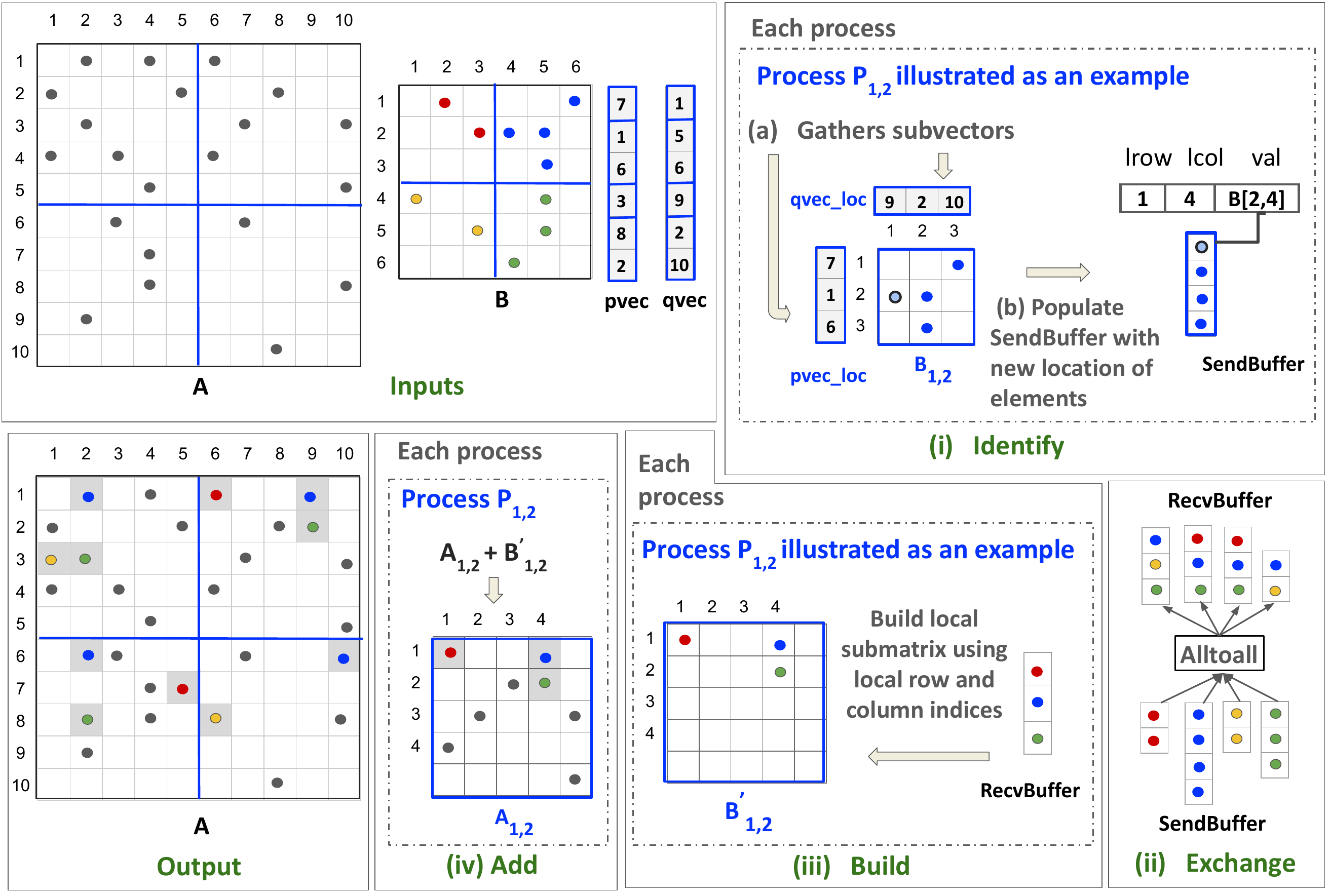}
\caption{This figure illustrates an example of the matrix assign operation. 
The inputs to the HipAssign algorithm are shown as matrix $\mat{B}$ (top left), which is to be assigned to a larger matrix $\mat{A}$ (top left) in corresponding locations indicated by the row indices vector $\vect{pvec}$ and column indices vector $\vect{qvec}$. 
HipAssign extends and permutes matrix $\mat{B}$ to form a new matrix $\mat{B}^\prime$ and then updates matrix $\mat{A}$ with the operation $\mat{A} = \mat{A} + \mat{B}^\prime$.
(i) In the first step of the algorithm, {\em Identify}, each process gathers the subvectors associated with its indices of the local submatrix of $\mat{B}$.
As an example in substep (a), process $P_{1,2}$ forms $\vect{pvec\_loc}$ with gathered subvectors from $P_{1,1}$ and $P_{1,2}$, and forms $\vect{qvec\_loc}$ with gathered subvectors from  $P_{2,1}$ and $P_{2,2}$. 
Then as shown in substep (b), the \texttt{SendBuffer} populated with the new locations of the local elements of $\mat{B}$, computed based on $\vect{pvec\_loc}$ and $\vect{qvec\_loc}$.
(ii) An \texttt{Alltoall} communication is executed to exchange elements between processes.
(iii) In {\em Build} step, each process constructs its local submatrix of $\mat{B}^\prime$  using the elements received in the \texttt{RecvBuffer}.
(iv) Each process then adds its local submatrix of $\mat{A}$ and $\mat{B}^\prime$. 
The addition of local submatrices $\mat{A}_{1,2}$ and $\mat{B}^\prime_{1,2}$ using a select-second operation is shown as an example in the figure (bottom left, highlighted in gray and colors).
The colored nonzeros with a gray background represent the nonzero elements assigned by the HipAssign operation.
$\mat{A}$ (bottom left) is the final output, showing the result of the assignment operation.}
\label{fig-Assign}
\end{figure}

The Identify step of HipAssign determines the new location of each element of $\mat{B}$ within the expanded and reordered matrix $\mat{B}^\prime$ and fills a send buffer for communication. This step has two substeps applied to the input matrix $\mat{B}$, as illustrated in Figure~\ref{fig-Assign}(i).

{\bf(a) Gather necessary indices.} 
The permutation and expansion of $\mat{B}$ follows $\mat{B}^\prime[\vect{pvec}[i]][\vect{qvec}[j]] = \mat{B}[i][j]$.
Unlike HipPerm, we do not need to swap indices and values of the vectors because they are already aligned in terms of indices. 

Hence, each process just gathers $\vect{pvec}$ along the process grid row and $\vect{qvec}$ along the process grid column to form local vectors $\vect{pvec\_loc}$ and $\vect{qvec\_loc}$, respectively (Line 2 in Algorithm ~\ref{algo:hipassign}). 
Two \texttt{Allgather} operations, each involving $\sqrt{p}$ processes, are required for this step.
In Figure~\ref {fig-Assign}(i), process $P_{1,2}$ is illustrated as an example in substep (a). 

{\bf(b) Identify destination process of local elements and populate a send buffer.} 
In substep (b) of {\em Identify} step, each process $P_{r,c}$ populates a \texttt{SendBuffer} using the inputs of local submatrix of $\mat{B}$ and local vectors $\vect{pvec\_loc}$ and $\vect{qvec\_loc}$ as shown in step (b) of Figure ~\ref{fig-Assign}(i).
Let $\mat{B}_{r,c}$ be the local submatrix on $P_{r,c}$, then for each nonzero element $\mat{B}_{r,c}[i][j]$, new location is computed as $\vect{pvec[i]}$ and $\vect{qvec[j]}$.
Since this new location may reside on a different process in a distributed system,
The destination process is computed as well.
As shown in step (b) of Figure~\ref {fig-Assign}(i), for each element, a triple $(lrow, lcol,val)$ is placed into the \texttt{SendBuffer} where $lrow$ and $lcol$ represent local row and column indices within the local submatrix in the destination process and $val$ is the nonzero value.
Similar to the discussion in HipPerm, the placement of elements in \texttt{SendBuffer} is according to the destination process due to the grouping of \texttt{SendBuffer} by the destination process id. 
The algorithm to populate the \texttt{SendBuffer} is described in Algorithm ~\ref{algo:threadComputation}.  

\subsubsection{Step 2: Exchange Data}
After all processes populate their \texttt{SendBuffer}, an exchange of data is performed using an \texttt{Alltoall} communication.

\subsubsection{Step 3: Build the Expanded and Permuted Matrix}
After the MPI \texttt{Alltoall} step, the received data is used to construct the local submatrices of $\mat{B}^\prime$ as illustrated in Figure ~\ref{fig-Assign}(iii). 
Unlike HipPerm, where the matrix dimensions remain unchanged at this step, $\mat{B}^\prime$ is built with dimensions $m\times n$, reflecting the target structure of $\mat{A}$. 
The local computation to build a local submatrix on each process is as explained in subsection ~\ref{sec:build-step}.

\subsubsection{Step 4: Add}
The final step adds matrix $\mat{B}^\prime$ to $\mat{A}$ in parallel with a user-defined function~\cite{hussain2022parallel}.
The result of HipAssign is the update matrix $\mat{A}$.

\begin{algorithm}[!t]
\caption{HipAssign: Assigning a distributed sparse matrix to a larger distributed sparse matrix}
\label{algo:hipassign}
\textbf{Input and Output:} Input sparse matrices $\mat{A} \in \mathbb{R}^{m \times n}$ and $\mat{B} \in \mathbb{R}^{m^\prime \times n^\prime}$  permutation vectors $\vect{pvec}$ and $\vect{qvec}$. Output updated matrix $\mat{A} \in \mathbb{R}^{m \times n}$. All matrices and vectors are distributed on a $\sqrt{p}\times \sqrt{p}$ process grid.
\begin{algorithmic}[1]
\Procedure{HipAssign}{$\mat{A}, \mat{B}, \vect{pvec}, \vect{qvec}$}
\State $\vect{pvec\_loc}, \vect{qvec\_loc} \gets$ Gather corresponding subvectors of $\vect{pvec}$ and $\vect{qvec}$  
\For{Every submatrix $\mat{B}_{r,c}$ in process $P_{r,c}$ in \textbf{parallel}} \Comment{MPI parallel}
\State \SendBuffer $\gets \Call{PrepareSendBuffer}{\mat{B}_{r,c},\vect{pvec\_loc}, \vect{qvec\_loc}}$
\EndFor
\State $\texttt{RecvBuffer} \gets \Call{MPI\_Alltoall}{\texttt{SendBuffer}}$ 
\For{Every process $P_{r,c}$ in \textbf{parallel}}
\State $\mat{B^\prime}_{r,c} \gets \Call{BuildLocalMatrix}{\texttt{RecvBuffer}}$ \Comment{$\mat{B}_{r,c} \in \mathbb{R}^{m\times n}$}
\State $\mat{A}_{r,c} \gets \mat{A}_{r,c} + \mat{B^\prime}_{r,c}$ \Comment{Addition operation is user-defined}
\EndFor
\State \Return $\mat{A}$
\EndProcedure
\end{algorithmic}
\end{algorithm}

\subsection{HipAssign Complexity Analysis}
To analyze the complexity of HipAssign, we consider each step of the algorithm separately.
To simplify the calculation, we assume both input matrices to be square, such as $\mat{A}\in \mathbb{R}^{n\times n}$ and $\mat{B}\in \mathbb{R}^{m\times m}$ where $m < n$ and lengths of the vectors are $|\vect{pvec}|=|\vect{qvec}|=m$.

\textbf{\em Identify: } 
In this step, two \texttt{Allgather} communication is applied to gather subvectors, each among $\sqrt{p}$ processes. Considering the Ring algorithm for \texttt{Allgather}, the communication complexity is $\mathcal{O}(\alpha (\sqrt{p}-1) + \beta\frac{m}{\sqrt{p}})$.
In the next substep, the computation on each local submatrix follows Algorithm ~\ref{algo:threadComputation}, we can refer to the complexity analysis of HipPerm ~\ref{sec:perm_analysis} for this step.
Let $nnz(\mat{B})$ be the number of nonzero elements of matrix $\mat{B}$, the computation complexity on each process is approximately $\mathcal{O}(\frac{nnz(\mat{B})}{p*\texttt{nthreads}})$.

\textbf{\em Exchange: } 
In the second step, the communication complexity of \texttt{Alltoall} with a pairwise exchange algorithm is $\mathcal{O}(\alpha (p-1)+ \beta \frac{nnz(\mat{B})}{p})$.

\textbf{\em Build: } 
Building matrix $\mat{B}^\prime$ locally on each process approximately takes $\mathcal{O}(\frac{nnz(\mat{B})}{p * \texttt{nthreads}})$.

\textbf{\em Add: } 
Adding two sparse matrices $\mat{B}^\prime$ and $\mat{A}$, considering two matrices are distributed evenly, require each process take $\mathcal{O}(\frac{nnz(\mat{A}) + nnz(\mat{B})}{p* \texttt{nthreads}})$ time complexity. 
\section{Results}
\subsection{Experimental Setup}
\paragraph{Experiment Objectives} 
We designed experiments to evaluate the performance of our algorithms in terms of runtime speedup, scalability, and overall efficiency. 
We aim to cover a range of matrix computation applications, each characterized by distinct input/output sparsity patterns.
For matrix permutation, we consider both random permutation and matrix reordering, demonstrating the robustness of our approach while highlighting that performance is sensitive to the specific application and underlying sparsity structure.
In submatrix extraction, we demonstrate the algorithm's applicability through applications involving both random submatrix selection and specific cluster extraction.
Furthermore, we integrate permutation and assignment operations to demonstrate their application in streaming batch updates for distributed graphs.

We demonstrate the scalability of our distributed algorithms at both the process and thread levels.
For large-scale matrices, we show scalability up to 4096 MPI processes (512 nodes; 65,536 cores).
Furthermore, we benchmark our methods against CombBLAS, a leading 2D distributed memory sparse library; PETSc, representing the 1D distributed memory approach; and GraphBLAS, a shared-memory baseline known for its superior single-node performance.

\paragraph{Implementation Details} 
Our algorithms are implemented in C/C++ and compiled using gcc 11.2.0.
We employed Cray’s MPI and OpenMP for hybrid parallelization. We used data structures from the CombBLAS library~\cite{azad2021combinatorial} that distribute vectors and matrices on a 2D square process grid.

\paragraph{Datasets}
The datasets used in our experiments are listed in Table~\ref{tab:datasets}.
We selected them to cover a diverse range of real-world and synthetic matrices and graphs, varying in size, structure, and sparsity.
For instance, our selection includes road networks with low average degrees, as well as social and protein networks with high average degrees.
We also used rectangular matrices and matrices arising from optimization problems. 
These matrices cover diverse sparsity patterns, ranging from diagonal or column-heavy nonzero patterns to more evenly distributed or entirely random graphs.
Additionally, we include extremely large matrices, such as \texttt{Hyperlink} and \texttt{Metaclust50}, to assess our approach at a large scale.

\begin{table}[!t]
    \centering
    \caption{The sparse matrices used in our experiments. All matrices are from the SuiteSparse matrix collection ~\cite{suitesparse}. Matrices are ordered according to their number of nonzeros (nnz).}
    \label{tab:datasets}
    \begin{tabular}{llrrrc}
        \toprule
        Name & Description & \#Rows  & \#nnz   & \#nnz/\#Rows
 & Symmetric \\ 
        \toprule    
        rail4284 & Linear Programming Problem & $4K\times 1.1M$& 11.2M & 2633 & Rectangular \\
        com-Orkut & Social Network  & $3.1M\times 3.1M$ & 234.3M & 76  & Yes  \\
        wikipedia-2007 & Hyperlink network & $3.5M\times 3.5M$ &  45M & 13 & No \\
        soc-LiveJournal1 &  Social network & $4.8M\times 4.8M$ & 68.9M & 14 & No \\
        Hardesty3 & Computer Vision Problem & $8.2M\times 7.6M$& 40.5M & 4.9 & Rectangular\\
        wb-edu & Web crawl & $9.8M\times 9.8M$ & 57M & 6 & No \\ 
        stokes & Semiconductor Process Problem & $11M\times 11M$ & 349M & 31 & No \\
        relat9 &  Combinatorial Problem & $12.3M\times 549K$& 39M & 3.15 & Rectangular \\
        uk-2002 & Web network & $18.5M\times 18.5M$ & 298M & 16 & No \\
        GAP-road & US road network & $23.9M\times 23.9M$ &  57M & 2 & Yes  \\
        nlpkkt240 & Optimization Problem & $27.9M\times 27.9M$ & 774M & 28 & Yes \\
        twitter7 & Social network  & $41.6M\times 41.6M$ & 1.5B & 35 & No \\
        GAP-web & Web crawl & $50.6M\times 50.6M$ & 1.9B & 38 & No \\
        europe-osm  & Europe road network  & $50.9M\times 50.9M$ & 108M & 2 & Yes \\
        com-Friendster & Social network & $65.6M\times 65.6M$ & 3.6B & 55 & Yes \\
        webbase & Web crawl  & $118M\times118M$ &  1.02B & 9 & No  \\
        GAP-urand & Random network & $134.2M\times 134.2M$ & 4.3B & 16 & Yes\\
        Metaclust50 & Protein network & $282M\times 282M$ & 42.8B & 151 & Yes \\
        Hyperlink & Web graph  & $3.44B\times 3.44B$ & 128B & 37 & No \\
        \bottomrule
    \end{tabular}
\end{table}

\paragraph{Computing Platforms}
We conducted most of our experiments on Big Red 200 at Indiana University and the Perlmutter supercomputer at NERSC, both HPE Cray EX systems.
Each compute node of Big Red 200 has 256 GB of memory and two 64-core 2.25 GHz AMD EPYC 7742 processes. Each compute node of the Perlmutter is equipped with two 64-core 2.25 GHz AMD EPYC 7763 CPUs and 512GB of memory.
In all experiments on Big Red 200 and Perlmutter, we used 8 MPI processes per node and 16 OpenMP threads per process.
We performed batch streaming update experiments on Grace supercomputer at Texas A\&M University, where each node has 384 GB of memory and two 24-core 3.0 GHz Intel Xeon 6248R processes. 
For consistency with our other experiments, we configured the Grace nodes to use 250 GB of memory.
For the experiments on Grace, we used 4 MPI processes per node and 12 OpenMP threads per process.

\subsection{Experiments with Matrix Permutations}
We evaluate matrix permutations in two scenarios: (a) random permutation and (b) matrix reordering.
In the first scenario, we randomly permute a distributed matrix to ensure that nonzero entries are uniformly distributed across processes. This permutation helps balance the load in subsequent graph and sparse matrix algorithms. 
The goal of random permutation is to destroy any structure present in the matrix. It is common in most distributed-memory libraries, including CombBLAS and PETSc.

In the second scenario, we apply a given permutation vector to reorder the matrix, concentrating nonzero entries near the diagonal. This approach reduces matrix bandwidth in iterative solvers or ensures a zero-free diagonal during the pivoting step of direct solvers. Various algorithms can be used to obtain such orderings, including Reverse Cuthill-McKee (RCM)~\cite{cuthill1969reducing, azad2017reverse}, minimum-degree ordering~\cite{george1989evolution,karypis1998parallel}, and bipartite matching~\cite{duff2001algorithms, azad2020AWPM}.
In this paper, we used the RCM ordering to permute matrices for concentrating nonzero entries near the diagonal.

\begin{table}[!t]
    \centering
     \caption{
     Performance comparison of HipPerm, CombBLAS, and PETSc in terms of execution time (seconds). All experiments except the last two rows were conducted on the Big Red 200 supercomputer using 8 nodes, with a total of 64 processes (16 threads/cores per process). Metaclust50 and Hyperlink permutations are performed on the Perlmutter supercomputer using 128 nodes, with a total of 1024 processes (16 threads per process). LI stands for the load imbalance ratio. RCM ordering does not apply for rectangular matrices as denoted by ``NA". 
     PETSc experiments cannot be performed for larger matrices due to file IO errors.
     Because of allocation limitations, we did not perform RCM reordering experiments for Hyperlink and Metaclust50.
     }
    \label{tab:permutation_comparison}
    \begin{tabular}{l r | r r r r | r r r}
    \toprule
     &  & \multicolumn{ 4}{c}{Random Permutation} & \multicolumn{ 3}{c}{RCM Reordering} \\
    Matrix & LI (before) & LI (after) & HipPerm & CombBLAS & PETSc & LI (after) &  HipPerm & CombBLAS \\
    \toprule
        rail4284 & 3.0 & 1.01 & 0.07 & 3.31 & 0.81 & NA & NA & NA\\
        com-Orkut & 5.9 & 1.01 & 1.82 & 80.21 & 2.58 & 9.8 & 4.67 & 138.7\\
        wikipedia-2007 & 11.9 & 1.04 & 0.66 & 31 & 8.2 & 16.9 & 2.2 & 55.3 \\
        soc-LiveJournal1 & 15.4 & 1.05 & 1.26 & 56.6 & 4.92 & 14.1 & 2.19 & 89.5 \\
        Hardesty3 & 7.5 & 1.00 & 0.42 & 24.2 & 1.34 & NA & NA & NA\\
        wb-edu  & 8.8 & 1.23 & 0.89 & 27.7 & 1.8  & 8.9 & 1.23 & 40.9 \\
        stokes  & 5 & 1.2 & 1.92 & 110.5 & 9.22 & 2.8 & 2.98 & 132.68\\
        relat9 & 7.3 & 1.00 & 0.24 & 27.3 & 1.28 & NA & NA & NA\\
        uk-2002 & 9.6 & 1.14 & 3.68 & 175.2 & 4.71 & 12.4 & 7.64 & 259.8\\
        GAP-road & 7.1 & 1.0 & 0.99 & 24.2 & 17.40 & 8.3 & 2.12 & 38.2\\
        nlpkkt240 & 7.6 & 1.2 & 5.41 & 348.1 & 15.74 & 7.9 & 12.87 & 448.1\\
        twitter7 & 9.3 & 1.03 & 18.48 & 831.5 & 277.95 & 10.8 & 42.23 & 1246.9 \\
        GAP-web & 8.7 & 1.02 & 20.9 & 1008.9 & 31.02 & 9.9 & 73.61 & 1638.2\\
        europe\_osm & 7.7 & 1.0 & 2.05 & 46.4 & 3.61 & 8.1 & 3.97 & 66.39\\
        com-Friendster & 9.0 & 1.0 & 32.05 & 1545.4 & IO ERR & 22.8 & 158.16 & 1575.8\\
        webbase & 9.2 & 1.2 & 13.62 & 663.2 & 12.98 & 9.3 & 133.91 & 917.13\\
        GAP-urand & 1 & 1.0 & 19.97 & 474.4 & IO ERR & 1.4 & 22.73 & 502.6\\
        \midrule
        Metaclust50 & 1.14 & 1.00 & 9.38 & 416.37 & IO ERR & - & - & -\\
        Hyperlink & 1.36  & - & 53.12  & - & IO ERR & - & - & -\\
    \bottomrule
    \end{tabular}
\end{table}

For random permutations, permutation vectors are generated randomly, while for RCM ordering, they are obtained from a distributed memory implementation of the RCM algorithm~\cite{azad2017reverse}.
We compare the performance of HipPerm CombBLAS and PETSc.
More specifically, we used the {\tt SubsRef\_SR} function in CombBLAS and the function {\tt MatPermute} in PETSc.
CombBLAS distributes matrices and vectors across a 2D process grid, whereas PETSc utilizes a 1D grid for data distribution.
We maintained consistent configurations for the number of nodes, MPI processes, and OpenMP threads across all performance comparisons.

\subsection{Performance of HipPerm relative to CombBLAS and PETSc}
Table~\ref{tab:permutation_comparison} shows the runtime of HipPerm, CombBLAS, PETSc on 64 processes of the Big Red 200 supercomputer. 
We also present the load imbalance before and after the permutations. In this context, load imbalance is measured as the ratio of the maximum non-zero entries (nnz) in a process to the average nnz per process. For example, a random matrix, GAP\_urand, has a load imbalance of 1 before permutation, where an imbalance factor of 1 indicates that all processes have an equal number of nonzeros in their local submatrix. In contrast, scale-free social networks like soc-LiveJournal1 exhibit a load imbalance of 15.5 before permutation.
Applying a random permutation to a matrix typically reduces the load imbalance factor, bringing it closer to 1. On the other hand, RCM reordering may increase the load imbalance factor as it moves nonzeros close to the diagonal.

\paragraph{Random permutation}
Table~\ref{tab:permutation_comparison} shows that HipPerm runs faster than CombBLAS and PETSc when we perform random permutations. 
We observe an average speedup of 55 (stdev = 108) over CombBLAS and 5.54 (stdev = 5.16) over PETSc permutation.
The lower performance of random permutations in CombBLAS stems from its implementation of symmetric permutations as two SpGEMM operations, $\mat{P}\mat{A}\mat{P}\transpose$. Its distributed SpGEMM relies on sparse SUMMA~\cite{bulucc2012parallel}, which keeps the output stationary while moving both input matrices.
In particular, sparse SUMMA broadcasts submatrices of $\mat{A}$ to remote processes that may not need them, substantially increasing communication overhead.
For several datasets, we observed that CombBLAS's communication time alone exceeded the total runtime of HipPerm. This result reinforces our argument that a general-purpose SpGEMM is too costly for distributed matrix permutation.

The last two rows in Table~\ref{tab:permutation_comparison} show a large-scale permutation experiment using 128 nodes (16,384 cores) on the Perlmutter supercomputer. At this extreme scale, HipPerm is more than $50\times$ faster than CombBLAS for Metaclust50.
The random permutation on Hyperlink could not be completed with CombBLAS due to an internal error.
PETSc also fails on these matrices, reporting a file I/O error.

Table~\ref{tab:permutation_comparison} shows that matrices with higher load imbalance factors take longer to permute. 
For example, GAP-urand (imbalance 1.0) runs faster than com-Friendster (imbalance 9.0) despite the former having more nonzeros. 
This indicates that the load imbalance increases both the computation and communication overhead of matrix permutation.

\paragraph{RCM reordering}
Table~\ref{tab:permutation_comparison} also reports the performance of HipPerm and CombBLAS with RCM orderings. 
As before, HipPerm is an order of magnitude faster than CombBLAS across all matrices. However, RCM reordering takes longer than random permutations for both libraries because it clusters nonzeros near the diagonal, increasing the imbalance factor after reordering. 
Consequently, diagonal processes receive more data and perform more work, raising both computation and communication costs and thus the overall runtime.

\begin{figure}[!t]
\centering
\begin{minipage}[b]{0.31\textwidth}
    \centering  
    \begin{tikzpicture}
    \begin{axis}[
    xlabel={Number of Processes},
    ylabel={Runtime (s)},
    ylabel style={yshift=-0.4em}, 
    align=left,
    legend style={at={(0.45,1.05)}, anchor=south, draw=none, legend columns=2, column sep=1ex},
    log basis x={2}, 
    log basis y={2}, 
    xmode=log, 
    ymode=log,
    xtick=data, 
    xticklabel style={rotate=45, anchor=east, yshift=-0.5em}, 
    ytick={ 1/4, 1, 4, 16, 64, 256, 1024, 4096},
    ymin = 0.1, ymax = 10000,
    grid=both, 
    legend cell align={left},
    scale only axis,
    scale=0.8,
    width=\linewidth,
    height=\linewidth
]
\definecolor{color1}{rgb}{0.0, 0.0, 1.0} 
\definecolor{color2}{rgb}{1.0, 0.0, 0.0} 
\definecolor{color3}{rgb}{0.0, 1.0, 0.0} 
\definecolor{color4}{rgb}{1.0, 0.6, 0.0} 
\definecolor{color5}{rgb}{0.5, 0.0, 1.0} 
\definecolor{color6}{rgb}{0.0, 0.7, 0.7} 
\definecolor{color7}{rgb}{0.5, 0.3, 0.0} 
\definecolor{color8}{rgb}{0.9, 0.1, 0.9} %
\addplot[dashed, thick, color=color1, forget plot, mark=square*, mark options={scale=0.5}] table [x index=0, y index=1] {\scalingdatatable}; 
\addplot[solid, thick, color=color1, forget plot,mark=square*, mark options={scale=0.5}] table [x index=0, y index=9] {\scalingdatatable}; 
\addlegendimage{only marks, mark=square*, mark options={scale=1, fill}, fill=color1}
\addlegendentry{Com-Orkut}

\addplot[dashed, thick, color=color4, forget plot, mark=diamond*, mark options={scale=0.5} ] table [x index=0, y index=4] {\scalingdatatable}; 
\addplot[solid, thick, color=color4, forget plot, mark=diamond*, mark options={scale=0.5} ] table [x index=0, y index=12] {\scalingdatatable}; 
\addlegendimage{only marks, mark=diamond*, mark options={scale=1, fill}, fill=color4}
\addlegendentry{wb-edu}

\addplot[dashed, thick, color=color2, forget plot, mark=triangle*, mark options={scale=0.5}] table [x index=0, y index=2] {\scalingdatatable}; 
\addplot[solid, thick, color=color2, forget plot, mark=triangle*, mark options={scale=0.5}] table [x index=0, y index=10] {\scalingdatatable}; 
\addlegendimage{only marks, mark=triangle*, mark options={scale=1, fill}, fill=color2}
\addlegendentry{wikipedia2007}

\addplot[dashed, thick, color=color5, forget plot, mark=x, mark options={scale=0.8}] table [x index=0, y index=5] {\scalingdatatable}; 
\addplot[solid, thick, color=color5, forget plot, mark=x, mark options={scale=0.8}] table [x index=0, y index=13] {\scalingdatatable}; 
\addlegendimage{only marks, mark=x, mark options={scale=1, fill}, fill=color5}
\addlegendentry{uk-2002}

\addplot[dashed, thick, color=color3, forget plot, mark=o, mark options={scale=0.5} ] table [x index=0, y index=3] {\scalingdatatable}; 
\addplot[solid, thick, color=color3, forget plot, mark=o, mark options={scale=0.5}] table [x index=0, y index=11] {\scalingdatatable}; 
\addlegendimage{only marks, mark=o, mark options={scale=1, fill}, fill=color3}
\addlegendentry{soc-LiveJournal1}

\addplot[dashed, thick, color=color7, forget plot, mark=oplus*, mark options={scale=0.5}] table [x index=0, y index=7] {\scalingdatatable}; 
\addplot[solid, thick, color=color7, forget plot, mark=oplus*, mark options={scale=0.5}] table [x index=0, y index=15] {\scalingdatatable}; 
\addlegendimage{only marks, mark=oplus*, mark options={scale=1, fill}, fill=color7, scale=1.2}
\addlegendentry{twitter7}

\addplot[dashed, thick, color=color6, forget plot, mark=pentagon*, mark options={scale=0.5}] table [x index=0, y index=6] {\scalingdatatable}; 
\addplot[solid, thick, color=color6, forget plot, mark=pentagon*, mark options={scale=0.5}] table [x index=0, y index=14] {\scalingdatatable}; 
\addlegendimage{only marks, mark=pentagon*, mark options={scale=1, fill}, fill=color6, scale=1.2}
\addlegendentry{europe\_osm}

\addplot[dashed, thick, color=color8, forget plot, mark=halfcircle*, mark options={scale=0.5}] table [x index=0, y index=8] {\scalingdatatable}; 
\addplot[solid, thick, color=color8, forget plot, mark=halfcircle*, mark options={scale=0.5}] table [x index=0, y index=16] {\scalingdatatable}; 
\addlegendimage{only marks, mark=halfcircle*, mark options={scale=1, fill}, fill=color8, scale=1.2}
\addlegendentry{GAP-road}
\end{axis}
    \end{tikzpicture}
    \subcaption{Smaller matrices on Big Red 200}\label{fig-scalability-small}
\end{minipage}%
\hfill
\begin{minipage}[b]{0.31\textwidth}
    \centering 
    \begin{tikzpicture}
    \begin{axis}[
    xlabel={Number of Processes},
    align=center,
    ylabel style={yshift=-0.4em}, 
    legend style={at={(0.5,1.05)}, anchor=south, draw=none, legend columns=2, column sep=1ex},
    log basis x={2}, 
    log basis y={2}, 
    xmode=log, 
    ymode=log,
    xtick=data, 
    xticklabel style={rotate=45, anchor=east, yshift=-0.5em}, 
    ytick={ 1/4, 1, 4, 16, 64, 256, 1024, 4096},
    ymin = 0.1, ymax = 7000,
    grid=both, 
    legend cell align={left},
    scale only axis,
    scale=0.8,
    width=\linewidth,
    height=\linewidth
]

\definecolor{color1}{rgb}{0.0, 0.0, 1.0} 
\definecolor{color2}{rgb}{1.0, 0.0, 0.0} 
\definecolor{color3}{rgb}{0.0, 1.0, 0.0} 
\definecolor{color4}{rgb}{1.0, 0.6, 0.0} 
\definecolor{color5}{rgb}{0.5, 0.0, 1.0} 

\addplot[dashed, thick, color=color2, forget plot, mark=square*, mark options={scale=0.5}] table [x index=0, y index=1] {\scalingdatalargetable}; 
\addplot[solid, thick, color=color2, forget plot, mark=square*, mark options={scale=0.5}] table [x index=0, y index=5] {\scalingdatalargetable}; 
\addlegendimage{only marks, mark=square*, mark options={scale=1, fill}, fill=color2}
\addlegendentry{GAP-web}

\addplot[dashed, thick, color=color3, forget plot, mark=x, mark options={scale=0.5}] table [x index=0, y index=2] {\scalingdatalargetable}; 
\addplot[solid, thick, color=color3, forget plot, mark=x, mark options={scale=0.5}] table [x index=0, y index=6] {\scalingdatalargetable}; 
\addlegendimage{only marks, mark=x, mark options={scale=1, fill}, fill=color3}
\addlegendentry{com-Friendster}

\addplot[dashed, thick, color=color4, forget plot, mark=triangle*, mark options={scale=0.5}] table [x index=0, y index=3] {\scalingdatalargetable}; 
\addplot[solid, thick, color=color4, forget plot, mark=triangle*, mark options={scale=0.5}] table [x index=0, y index=7] {\scalingdatalargetable}; 
\addlegendimage{only marks, mark=triangle*, mark options={scale=1, fill}, fill=color4}
\addlegendentry{webbase}

\addplot[dashed, thick, color=color5, forget plot, mark=diamond*, mark options={scale=0.5}] table [x index=0, y index=4] {\scalingdatalargetable}; 
\addplot[solid, thick, color=color5, forget plot, mark=diamond*, mark options={scale=0.5}] table [x index=0, y index=8] {\scalingdatalargetable}; 
\addlegendimage{only marks, mark=diamond*, mark options={scale=1, fill}, fill=color5}
\addlegendentry{GAP-urand}
\end{axis}
\end{tikzpicture}
\subcaption{Medium matrices on Big Red 200}\label{fig-scalability-medium}
\end{minipage}%
\hfill
\begin{minipage}[b]{0.31\textwidth}
\centering 
    \begin{tikzpicture}
    \begin{axis}[
    xlabel={Number of Processes},
    align=center,
    ylabel style={yshift=-0.4em}, 
    legend style={at={(0.5,1.05)}, anchor=south, draw=none, legend columns=1, column sep=1ex},
    log basis x={2}, 
    log basis y={2}, 
    xmode=log, 
    ymode=log,
    xtick=data, 
    xticklabel style={rotate=45, anchor=east, yshift=-0.5em}, 
    grid=both, 
    legend cell align={left},
    scale only axis,
    scale=0.8,
    width=\linewidth,
    height=\linewidth
]

\definecolor{color1}{rgb}{0.0, 0.0, 1.0} 
\definecolor{color2}{rgb}{1.0, 0.0, 0.0} 

\addplot[solid, thick, color=color2, forget plot, mark=square*, mark options={scale=0.5}] table [x index=0, y index=1] {\scalingLarge}; 
\addplot[dashed, thick, color=color2, forget plot, mark=square*, mark options={scale=0.5}] table [x index=0, y index=3] {\scalingLarge}; 
\addlegendimage{only marks, mark=square*, mark options={scale=1, fill}, fill=color2}
\addlegendentry{MetaClust}

\addplot[solid, thick, color=color1, forget plot, mark=diamond*, mark options={scale=0.5}] table [x index=0, y index=2] {\scalingLarge}; 
\addplot[dashed, thick, color=color1, forget plot, mark=diamond*, mark options={scale=0.5}] table [x index=0, y index=4] {\scalingLarge}; 
\addlegendimage{only marks, mark=diamond*, mark options={scale=1, fill}, fill=color1}
\addlegendentry{Hyperlink}
\end{axis}
\end{tikzpicture}
\subcaption{Large matrices on Perlmutter}
\label{fig-scalability-large} 

\end{minipage}%

\caption{Scalability of random permutations with HipPerm (solid lines) and CombBLAS (dashed lines) for different matrices. The y-axis denotes runtime in seconds. Both the x-axis and y-axis are shown on a logarithmic scale. Experiments in subfigures (a) and (b) were conducted on Big Red 200, while the experiment in subfigure (c) was conducted on Perlmutter. 
}
\label{fig-randp-scalability}
\end{figure}
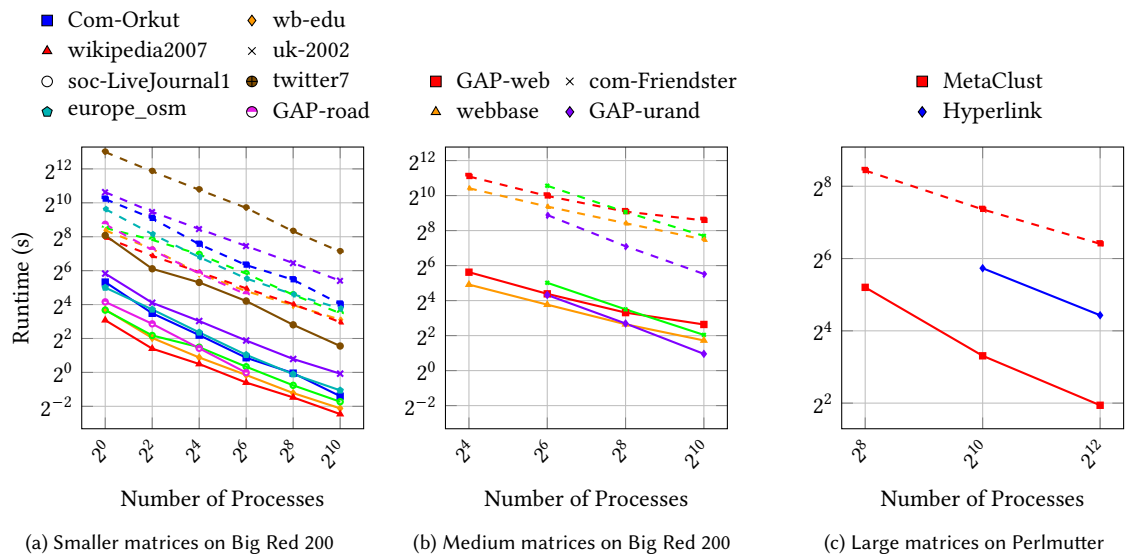

\subsection{Scalability of Matrix Permutation}
\paragraph{Random permutation}
Figure~\ref{fig-randp-scalability} illustrates the scalability of random permutations across three sets of graphs: small (Figure~\ref{fig-scalability-small}), medium (Figure~\ref{fig-scalability-medium}), and large (Figure~\ref{fig-scalability-large}) datasets. 
In all cases, the number of threads is fixed (16 threads per process), and the number of processes is increased. 
Both HipPerm and CombBLAS scale to thousands of processes, but HipPerm remains orders of magnitude faster in all cases. 
As shown in Figure~\ref{fig-scalability-small}, HipPerm achieves an average 60$\times$ speedup when scaling from 1 to 1,024 processes. Figure~\ref{fig-scalability-large} further demonstrates efficient scaling up to 4,096 processes (65,536 cores) on Perlmutter. 
CombBLAS failed to permute the Hyperlink dataset due to an internal error, and at this scale HipPerm is about an order of magnitude faster than CombBLAS’s random permutation.

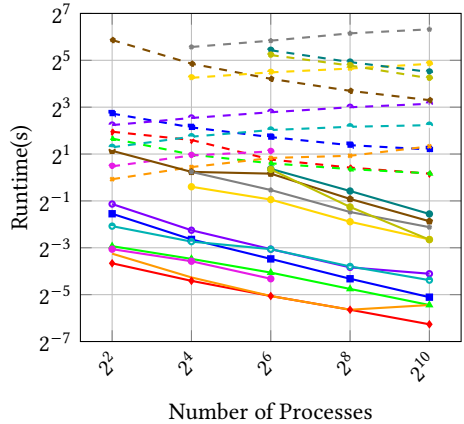
\begin{figure}[!t]
\centering
\begin{tikzpicture}
    \begin{axis}[
    xlabel={Number of Processes},
    ylabel={Runtime(s)},
    ylabel style={yshift=-0.4em}, 
    align=center,
    log basis x={2}, 
    log basis y={2}, 
    xmode=log, 
    ymode=log,
    xtick={4, 16, 64, 256, 1024}, 
    xticklabel style={rotate=45, anchor=east, yshift=-0.5em}, 
     legend style={at={(0.45,1.05)}, anchor=south, draw=none, legend columns=6, column sep=1ex},
    ytick={0.0078125, 0.03125, 0.125, 0.5, 2, 8, 32,128},
    ymin=0.0078125,
    ymax=128,
    grid=both, 
    legend cell align={left},
    scale=0.6,
  scale only axis,
]
\definecolor{color1}{rgb}{0.0, 0.0, 1.0} 
\definecolor{color2}{rgb}{1.0, 0.0, 0.0} 
\definecolor{color3}{rgb}{0.0, 1.0, 0.0} 
\definecolor{color4}{rgb}{1.0, 0.6, 0.0} 
\definecolor{color5}{rgb}{0.5, 0.0, 1.0} 
\definecolor{color6}{rgb}{0.0, 0.7, 0.7} 
\definecolor{color7}{rgb}{0.5, 0.3, 0.0} 
\definecolor{color8}{rgb}{0.9, 0.1, 0.9} 
\definecolor{color9}{rgb}{0.5, 0.5, 0.5} 
\definecolor{color10}{rgb}{1.0, 0.84, 0.0} 
\definecolor{color11}{rgb}{0.0, 0.5, 0.5} 
\definecolor{color12}{rgb}{0.75, 0.75, 0.0} 
\addplot[dashed, thick, color=color1, forget plot, mark=square*, mark options={scale=0.5}] table [x index=0, y index=1] {\scalingcommunication}; 
\addplot[solid, thick, color=color1, forget plot, mark=square*, mark options={scale=0.5}] table [x index=0, y index=13] {\scalingcommunication}; 
\addlegendimage{only marks, mark=square*, mark options={scale=1, fill}, fill=color1}
\addlegendentry{Com-Orkut}
\addplot[dashed, thick, color=color2, forget plot, mark=diamond*, mark options={scale=0.5}] table [x index=0, y index=2] {\scalingcommunication}; 
\addplot[solid, thick, color=color2, forget plot, mark=diamond*, mark options={scale=0.5}] table [x index=0, y index=14] {\scalingcommunication}; 
\addlegendimage{only marks, mark=diamond*, mark options={scale=1, fill}, fill=color2}
\addlegendentry{wikipedia2007}
\addplot[dashed, thick, color=color3, forget plot, mark=triangle*, mark options={scale=0.5}] table [x index=0, y index=3] {\scalingcommunication}; 
\addplot[solid, thick, color=color3, forget plot, mark=triangle*, mark options={scale=0.5}] table [x index=0, y index=15] {\scalingcommunication}; 
\addlegendimage{only marks, mark=triangle*, mark options={scale=1, fill}, fill=color3}
\addlegendentry{soc-LiveJournal1}
\addplot[dashed, thick, color=color4, forget plot, mark=x, mark options={scale=0.5}] table [x index=0, y index=4] {\scalingcommunication}; 
\addplot[solid, thick, color=color4, forget plot, mark=x*, mark options={scale=0.5}] table [x index=0, y index=16] {\scalingcommunication}; 
\addlegendimage{only marks, mark=x, mark options={scale=1, fill}, fill=color4}
\addlegendentry{wb-edu}
\addplot[dashed, thick, color=color5, forget plot, mark=o, mark options={scale=0.5}] table [x index=0, y index=5] {\scalingcommunication}; 
\addplot[solid, thick, color=color5, forget plot, mark=o, mark options={scale=0.5}] table [x index=0, y index=17] {\scalingcommunication}; 
\addlegendimage{only marks, mark=o, mark options={scale=1, fill}, fill=color5}
\addlegendentry{uk-2002}
\addplot[dashed, thick, color=color8, forget plot, mark=oplus*, mark options={scale=0.5}] table [x index=0, y index=6] {\scalingcommunication}; 
\addplot[solid, thick, color=color8, forget plot, mark=oplus*, mark options={scale=0.5}] table [x index=0, y index=18] {\scalingcommunication}; 
\addlegendimage{only marks, mark=oplus*, mark options={scale=1, fill}, fill=color8, scale=1.2}
\addlegendentry{GAP-road}
\addplot[dashed, thick, color=color7, forget plot, mark=pentagon*, mark options={scale=0.5}] table [x index=0, y index=7] {\scalingcommunication}; 
\addplot[solid, thick, color=color7, forget plot, mark=pentagon*, mark options={scale=0.5}] table [x index=0, y index=19] {\scalingcommunication}; 
\addlegendimage{only marks, mark=pentagon*, mark options={scale=1, fill}, fill=color7, scale=1.2}
\addlegendentry{twitter7}
\addplot[dashed, thick, color=color6, forget plot, mark=halfcircle*, mark options={scale=0.5}] table [x index=0, y index=8] {\scalingcommunication}; 
\addplot[solid, thick, color=color6, forget plot, mark=halfcircle*, mark options={scale=0.5}] table [x index=0, y index=20] {\scalingcommunication}; 
\addlegendimage{only marks, mark=halfcircle*, mark options={scale=1, fill}, fill=color6, scale=1.2}
\addlegendentry{europe\_osm}
\addplot[dashed, thick, color=color9, forget plot, mark=star, mark options={scale=0.5}] table [x index=0, y index=9] {\scalingcommunication}; 
\addplot[solid, thick, color=color9, forget plot, mark=star, mark options={scale=0.5}] table [x index=0, y index=21] {\scalingcommunication}; 
\addlegendimage{only marks, mark=star, mark options={scale=1, fill}, fill=color9, scale=1.2}
\addlegendentry{GAP-Web}
\addplot[dashed, thick, color=color10, forget plot, mark=otimes, mark options={scale=0.5}] table [x index=0, y index=10] {\scalingcommunication}; 
\addplot[solid, thick, color=color10, forget plot, mark=otimes, mark options={scale=0.5}] table [x index=0, y index=22] {\scalingcommunication}; 
\addlegendimage{only marks, mark=otimes, mark options={scale=1, fill}, fill=color10, scale=1.2}
\addlegendentry{webbase}
\addplot[dashed, thick, color=color11, forget plot, mark=*, mark options={scale=0.5}] table [x index=0, y index=11] {\scalingcommunication}; 
\addplot[solid, thick, color=color11, forget plot, mark=*, mark options={scale=0.5}] table [x index=0, y index=23] {\scalingcommunication}; 
\addlegendimage{only marks, mark=*, mark options={scale=1, fill}, fill=color11, scale=1.2}
\addlegendentry{com-Friendster}
\addplot[dashed, thick, color=color12, forget plot, mark=otimes*, mark options={scale=0.5}] table [x index=0, y index=12] {\scalingcommunication}; 
\addplot[solid, thick, color=color12, forget plot, mark=otimes*, mark options={scale=0.5}] table [x index=0, y index=24] {\scalingcommunication}; 
\addlegendimage{only marks, mark=otimes*, mark options={scale=1, fill}, fill=color12, scale=1.2}
\addlegendentry{GAP-urand}
\end{axis}
\end{tikzpicture}
\caption{Scalability of the communication time for HipPerm and CombBLAS permutations. Experiments were run on Big Red 200.}
\label{fig-scalability-communication}
\end{figure}
Figure~\ref{fig-scalability-communication} illustrates the communication scaling of both HipPerm and CombBLAS.
HipPerm’s communication scales much better than CombBLAS for all matrices. 
In particular, CombBLAS's communication does not scale effectively for matrices with special structures.
This limitation arises because CombBLAS relies on row- and column-wise broadcasts, which scale poorly for column- or diagonal-heavy matrices.

\begin{figure}[htbp]
\centering
\begin{tikzpicture}
    \begin{axis}[
    xlabel={Number of Processes},
    ylabel={Runtime(s)},
    align=center,
    ylabel style={yshift=-0.2em}, 
    legend style={at={(0.45,1.05)}, anchor=south, draw=none, legend columns=6, column sep=1ex},
    log basis x={2}, 
    log basis y={2}, 
    xmode=log, 
    ymode=log,
    xtick=data, 
    ytick={1, 4, 16, 64, 256, 1024, 4096},
    xticklabel style={rotate=45, anchor=east, yshift=-0.5em}, 
    grid=both, 
    legend cell align={left},
    scale=0.45,
    scale only axis,
]

\definecolor{color1}{rgb}{0.0, 0.0, 1.0} 
\definecolor{color2}{rgb}{1.0, 0.0, 0.0} 
\definecolor{color3}{rgb}{0.0, 1.0, 0.0} 
\definecolor{color4}{rgb}{1.0, 0.6, 0.0} 
\definecolor{color5}{rgb}{0.5, 0.0, 1.0} 
\definecolor{color6}{rgb}{0.5, 0.3, 0.0} 

\addplot[dashed, thick, color=color1, forget plot, mark=square*, mark options={scale=0.5}] table [x index=0, y index=1] {\scalingReordering}; 
\addplot[solid, thick, color=color1, forget plot, mark=square*, mark options={scale=0.5}] table [x index=0, y index=7] {\scalingReordering}; 
\addlegendimage{only marks, mark=square*, mark options={scale=1, fill}, fill=color1}
\addlegendentry{wikipedia-2007}

\addplot[dashed, thick, color=color2, forget plot, mark=diamond*, mark options={scale=0.5}] table [x index=0, y index=2] {\scalingReordering}; 
\addplot[solid, thick, color=color2, forget plot, mark=diamond*, mark options={scale=0.5}] table [x index=0, y index=8] {\scalingReordering}; 
\addlegendimage{only marks, mark=diamond*, mark options={scale=1, fill}, fill=color2}
\addlegendentry{soc-LiveJournal1}

\addplot[dashed, thick, color=color3, forget plot, mark=triangle*, mark options={scale=0.5}] table [x index=0, y index=3] {\scalingReordering}; 
\addplot[solid, thick, color=color3, forget plot, mark=triangle*, mark options={scale=0.5}] table [x index=0, y index=9] {\scalingReordering}; 
\addlegendimage{only marks, mark=triangle*, mark options={scale=1, fill}, fill=color3}
\addlegendentry{twitter7}

\addplot[dashed, thick, color=color4, forget plot, mark=oplus*, mark options={scale=0.75}] table [x index=0, y index=4] {\scalingReordering}; 
\addplot[solid, thick, color=color4, forget plot, mark=oplus*, mark options={scale=0.75}] table [x index=0, y index=10] {\scalingReordering}; 
\addlegendimage{only marks, mark=oplus*, mark options={scale=1, fill}, fill=color4}
\addlegendentry{com-Friendster}

\addplot[dashed, thick, color=color5, forget plot, mark=pentagon*, mark options={scale=0.75}] table [x index=0, y index=5] {\scalingReordering}; 
\addplot[solid, thick, color=color5, forget plot, mark=pentagon*, mark options={scale=0.75}] table [x index=0, y index=11] {\scalingReordering}; 
\addlegendimage{only marks, mark=pentagon*, mark options={scale=1, fill}, fill=color5}
\addlegendentry{GAP-urand}

\addplot[dashed, thick, color=color6, forget plot, mark=x, mark options={scale=1}] table [x index=0, y index=6] {\scalingReordering}; 
\addplot[solid, thick, color=color6, forget plot, mark=x, mark options={scale=1}] table [x index=0, y index=12] {\scalingReordering}; 
\addlegendimage{only marks, mark=x, mark options={scale=1, fill}, fill=color6}
\addlegendentry{nlpkkt240}

\end{axis}
\end{tikzpicture}
\caption{Scalability of HipPerm and CombBLAS when reordering matrices with RCM ordering on Big Red 200. Similar to other experiments, 16 threads per processor are used.}
\label{fig-reordering-scalability}
\end{figure}
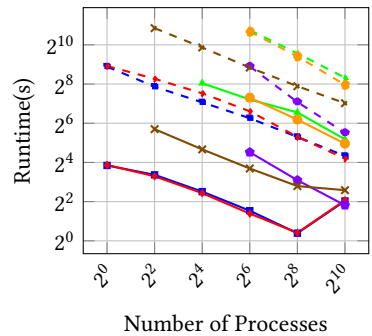
\paragraph{Scalability of RCM reordering}
Figure~\ref{fig-reordering-scalability} illustrates the scalability of HipPerm and CombBLAS permutations when reordering matrices using RCM ordering. 
The solid lines represent HipPerm, while the dashed lines represent CombBLAS permutations.
Similar to random permutations, both algorithms demonstrate good scalability; however, HipPerm remains orders of magnitude faster than CombBLAS across all experiments.
As shown in Table~\ref{tab:permutation_comparison}, RCM reordering often leads to a significant load imbalance in the output matrix (e.g., the imbalance factor reaches 18 for {\tt com-Friendster} after RCM reordering).
For such matrices, RCM reordering communication does not scale as effectively as random permutations, causing scalability to stall beyond 256 processes, as illustrated in Figure~\ref{fig-reordering-scalability}.

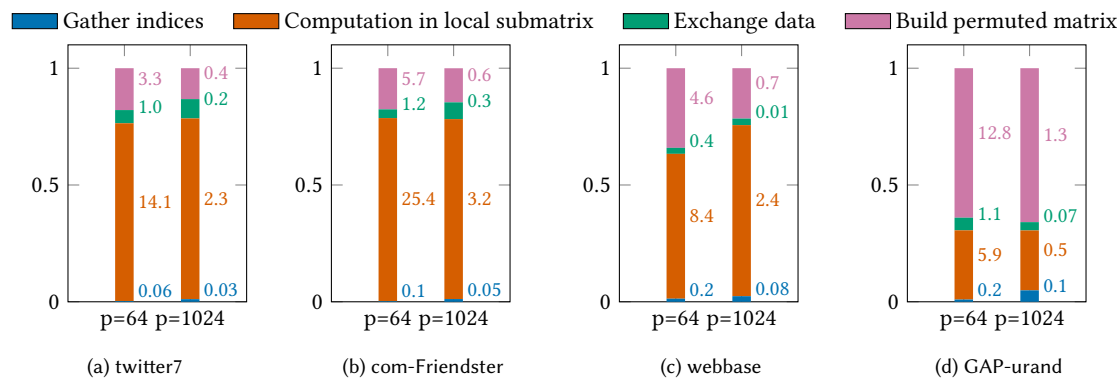
\begin{figure*}[htbp]
\centering
\begin{tikzpicture}
\begin{axis}[
  hide axis,
  width=5cm,
  height=5cm,
  xmin=0, xmax=1, ymin=0, ymax=1,
  legend columns=4,
  legend style={
    draw=none,
    at={(0.5,1.2)}, 
    anchor=south,
    /tikz/every even column/.append style={column sep=1em}
  },
  clip=false
]
\addlegendimage{area legend, ybar, draw=none, fill=oiBlue}
\addlegendentry{Gather indices}
\addlegendimage{area legend, ybar, draw=none, fill=oiRed}
\addlegendentry{Computation in local submatrix}
\addlegendimage{area legend, ybar, draw=none, fill=oiGreen}
\addlegendentry{Exchange data}
\addlegendimage{area legend, ybar, draw=none, fill=oiPink}
\addlegendentry{Build permuted matrix}
\end{axis}
\end{tikzpicture}
\begin{minipage}{0.24\textwidth}
\centering 
\begin{tikzpicture}
\begin{axis}[
    ybar stacked,
    bar width=7pt,
    width=5cm,
    height=5cm,
    xtick={0.8, 1.5},
    xticklabels={p=64, p=1024},
    ymin=0,
    nodes near coords,
    point meta=explicit symbolic, 
    every node near coord/.append style={font=\small, anchor=west, xshift=2pt, yshift=4pt},
    enlarge x limits={abs=0.75cm},
    x=1.25cm,
]
\addplot+[ybar, fill=oiBlue, draw=none, every node near coord/.append style={text=oiBlue}] coordinates {(0.8, 0.00325) [0.06] (1.5, 0.0112) [0.03]};
\addplot+[ybar, fill=oiRed, draw=none, every node near coord/.append style={text=oiRed}] coordinates {(0.8, 0.7610) [14.1] (1.5, 0.7738) [2.3]};
\addplot+[ybar, fill=oiGreen, draw=none, every node near coord/.append style={text=oiGreen}] coordinates {(0.8, 0.0565) [1.0] (1.5, 0.0827) [0.2]};
\addplot+[ybar, fill=oiPink, draw=none, every node near coord/.append style={text=oiPink}] coordinates {(0.8, 0.1793) [3.3] (1.5, 0.1323) [0.4]};
\end{axis}
\end{tikzpicture}
\subcaption{twitter7}
\end{minipage}%
\hfill
\begin{minipage}{0.24\textwidth}
\begin{tikzpicture}
\begin{axis}[
    ybar stacked,
    bar width=7pt,
    width=5cm,
    height=5cm,
    xtick={0.8, 1.5},
    xticklabels={p=64, p=1024},
    legend style={at={(0.5,-0.15)},
    anchor=north,legend columns=-1},
    ymin=0,
    nodes near coords,
    point meta=explicit symbolic, 
    every node near coord/.append style={font=\small, anchor=west, xshift=2pt, yshift=4pt}, 
    enlarge x limits={abs=0.75cm},
    x=1.25cm,
]
\addplot+[ybar,fill=oiBlue, draw=none, every node near coord/.append style={text=oiBlue}] coordinates {(0.8, 0.00278) [0.1] (1.5, 0.01168) [0.05]};
\addplot+[ybar,fill=oiRed, draw=none, every node near coord/.append style={text=oiRed}] coordinates {(0.8, 0.78364) [25.4] (1.5, 0.77032) [3.2]};
\addplot+[ybar,fill=oiGreen, draw=none, every node near coord/.append style={text=oiGreen}] coordinates {(0.8, 0.03735) [1.2] (1.5, 0.07175) [0.3]};
\addplot+[ybar,fill=oiPink, draw=none, every node near coord/.append style={text=oiPink}] coordinates {(0.8, 0.17623) [5.7] (1.5, 0.14597) [0.6]};
\end{axis}
\end{tikzpicture}
\subcaption{com-Friendster}
\end{minipage}%
\hfill
\begin{minipage}{0.24\textwidth}
\begin{tikzpicture}
\begin{axis}[
    ybar stacked,
    bar width=7pt,
    width=5cm,
    height=5cm,
    xtick={0.8, 1.5},
    xticklabels={p=64, p=1024},
    legend style={at={(0.5,-0.15)},
      anchor=north,legend columns=-1},
    ymin=0,
    nodes near coords,
    point meta=explicit symbolic, 
    every node near coord/.append style={font=\small, anchor=west, xshift=2pt, yshift=4pt}, 
    enlarge x limits={abs=0.75cm},
    x=1.25cm,
]
\addplot+[ybar,fill=oiBlue, draw=none, every node near coord/.append style={text=oiBlue}] coordinates {(0.8, 0.0132)[0.2] (1.5, 0.0239)[0.08]};
\addplot+[ybar,fill=oiRed, draw=none, every node near coord/.append style={text=oiRed}] coordinates {(0.8, 0.6201)[8.4] (1.5, 0.7324)[2.4]};
\addplot+[ybar,fill=oiGreen, draw=none, every node near coord/.append style={text=oiGreen}] coordinates {(0.8, 0.0257)[0.4] (1.5, 0.0279)[0.01]};
\addplot+[ybar,fill=oiPink, draw=none, every node near coord/.append style={text=oiPink}] coordinates {(0.8, 0.3409)[4.6] (1.5, 0.2158)[0.7]};
\end{axis}
\end{tikzpicture}
\subcaption{webbase}
\end{minipage}%
\hfill
\begin{minipage}{0.24\textwidth}
\begin{tikzpicture}
\begin{axis}[
    ybar stacked,
    bar width=7pt,
    width=5cm,
    height=5cm,
    xtick={0.8, 1.5},
    xticklabels={p=64, p=1024},
    legend style={at={(0.5,-0.15)},anchor=north,legend columns=-1},
    ymin=0,
    nodes near coords,
    point meta=explicit symbolic, 
    every node near coord/.append style={font=\small, anchor=west, xshift=2pt, yshift=4pt}, 
    enlarge x limits={abs=0.75cm},
    x=1.25cm,
]
\addplot+[ybar,fill=oiBlue, draw=none, every node near coord/.append style={text=oiBlue}] coordinates {(0.8, 0.0095)[0.2] (1.5, 0.0489)[0.1]};
\addplot+[ybar,fill=oiRed, draw=none, every node near coord/.append style={text=oiRed}] coordinates {(0.8, 0.2963)[5.9] (1.5, 0.2565)[0.5]};
\addplot+[ybar,fill=oiGreen, draw=none, every node near coord/.append style={text=oiGreen}] coordinates {(0.8, 0.0541)[1.1] (1.5, 0.0355)[0.07]};
\addplot+[ybar,fill=oiPink, draw=none, every node near coord/.append style={text=oiPink}] coordinates {(0.8, 0.6401)[12.8] (1.5, 0.6591)[1.3]};
\end{axis}
\end{tikzpicture}
\subcaption{GAP-urand}
\end{minipage}
\caption{Runtime breakdown (as a fraction of the total time) for random symmetric permutation of matrices using HipPerm on Big Red 200 with 64 and 1024 processes. The values next to each bar segment indicate the absolute runtime in seconds for each component. }
\label{fig-breakdown-permutation}
\end{figure*}
\subsection{Runtime Breakdown of HipPerm}
Figure ~\ref{fig-breakdown-permutation} presents the runtime breakdown of HipPerm with 64 processes (8 nodes) and 1024 processes (128 nodes) on Big Red 200.
The runtime is divided into four components: {\em Gather necessary indices}, {\em Computation in local matrix}, {\em Exchange data}, and {\em Build permuted matrix}. 
We observe that  {\em Computation in local matrix} and {\em Build permuted matrix} dominate the runtime, together accounting for more than 80\% of the total execution time.

The cost of each step and the scalability of each component depend on the matrix structure before and after permutation.
For {\em Computation in local matrix}, an imbalanced input matrix causes certain processes to handle a disproportionately large number of nonzeros, which leads to longer computation times.
Additionally, {\em Exchange data} step remains a minor contributor to overall runtime, even as the number of processes increases. 
For highly imbalanced input matrices, {\em Computation in local matrix} and {\em Exchange data} do not scale linearly because increasing the number of processes does not necessarily reduce the nonzeros per process.
Similarly, if the permutation reorders the matrix into a more imbalanced structure, the {\em Build permuted matrix} step can take a larger share of the runtime, and its scalability will be affected.

\begin{table}[!t]
\centering
\caption{Performance comparison of our permutation algorithm using 64 processes with 2 threads per process, with GraphBLAS and PETSc, both using 128 threads on a single node.
}
\label{tab:performance-graphblas}
\pgfplotstabletypeset[
    col sep=comma,
    string type, 
    every head row/.style={before row=\toprule, after row=\midrule},
    every last row/.style={after row=\bottomrule},
    columns={Dataset, GraphBLAS, OurPerm3, Petsc},  
    display columns/0/.style={column name=Graph},  
    display columns/1/.style={column name=\makecell{GraphBLAS(s)}},   
    display columns/2/.style={column name=\makecell{HipPerm(s)}},
    display columns/3/.style={column name=\makecell{PETSc(s)}}
]{\PermGraphBLASOurs}
\end{table}

\subsection{Shared-memory Performance and Thread Scalability}
While the previous analysis compared HipPerm with distributed-memory frameworks, this section evaluates its single-node performance. 
We compare HipPerm against SuiteSparse GraphBLAS~\cite{davis2023algorithm} and PETSc, as shown in Table~\ref{tab:performance-graphblas}. Since HipPerm is not designed as a shared-memory algorithm, we restrict the test to a small set of medium-scale matrices. In this experiment, HipPerm runs with 64 processes and 2 threads per process on a single node, matching the 128-core configuration of GraphBLAS and PETSc.
On a single node, our algorithm is 5.8$\times$ faster than PETSc but on average 3.8$\times$ slower than GraphBLAS. While 
GraphBLAS is highly optimized for shared-memory systems, HipPerm is designed for distributed-memory environments. Therefore, we do not recommend using HipPerm on a single node, where GraphBLAS or similar shared-memory libraries are more suitable.

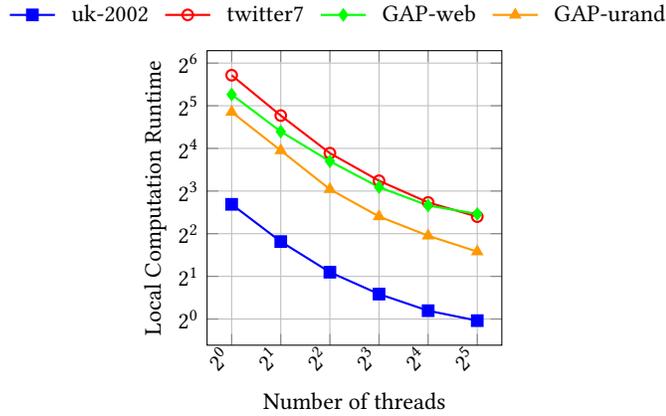
\begin{figure}[!t]
\centering
\begin{tikzpicture}
    \begin{axis}[
        xlabel={Number of threads},
        ylabel={Local Computation Runtime},
        ylabel style={yshift=-.2em},
        legend style={at={(0.45,1.05)}, anchor=south, draw=none, legend columns=4, column sep=1ex},
        log basis x={2}, 
        log basis y={2}, 
        xmode=log, 
        ymode=log,
        xtick=data, 
        ytick={1, 2, 4, 8, 16, 32, 64},
        xticklabel style={rotate=45, anchor=east},
        grid=both, 
        legend cell align={left},
        width=5.5cm, 
        height=5.5cm, 
    ]
    \definecolor{color1}{rgb}{0.0, 0.0, 1.0} 
    \definecolor{color2}{rgb}{1.0, 0.0, 0.0} 
    \definecolor{color3}{rgb}{0.0, 1.0, 0.0} 
    \definecolor{color4}{rgb}{1.0, 0.6, 0.0} 

    \addplot[solid, thick, color=color1, mark=square*, mark options={scale=1}] table [x index=0, y index=1] {\threadscalingdatatable}; 
    \addlegendentry{uk-2002}
    
    \addplot[solid, thick, color=color2, mark=o, mark options={scale=1}] table [x index=0, y index=2] {\threadscalingdatatable}; 
    \addlegendentry{twitter7}

    \addplot[solid, thick, color=color3, mark=diamond*, mark options={scale=1}] table [x index=0, y index=3] {\threadscalingdatatable}; 
    \addlegendentry{GAP-web}

    \addplot[solid, thick, color=color4, mark=triangle*, mark options={scale=1}] table [x index=0, y index=4] {\threadscalingdatatable}; 
    \addlegendentry{GAP-urand}
    \end{axis}
\end{tikzpicture}
\caption{Thread scalability of local computation consist of two steps(\textsc{PrepareSendBuffer} and \textsc{BuildLocalMatrix}) of HipPerm within a single process.}
\label{fig-thread-scalability}
\end{figure}

Next, we study the thread scalability of the local computations within a single process. 
Figure~\ref{fig-thread-scalability} shows that the local computation scales well up to 64 threads.
This scalability is achieved by the multithreaded computation in \textsc{PrepareSendBuffer} and \textsc{BuildLocalMatrix} steps, which ensures an even distribution of computational load across threads within each process and eliminate thread synchronization entirely.
The results suggest that local computation benefits from multithreading, though the speedup decreases at higher thread counts due to overheads.

\begin{figure}[htbp]
\centering
\begin{tikzpicture}
\begin{axis}[
    ybar stacked,
    hide axis,
    xmin=0,
    xmax=1,
    ymin=0,
    ymax=1,
    legend style={
        draw=none, 
        at={(-1,0.3)}, 
        anchor=north,
        legend columns=5, 
        /tikz/every even column/.append style={column sep=1em}
    }
]
\definecolor{color1}{rgb}{0.0, 0.0, 1.0} 
\definecolor{color2}{rgb}{1.0, 0.0, 0.0} 
\definecolor{color3}{rgb}{0.0, 1.0, 0.0} 
\definecolor{color4}{rgb}{1.0, 0.6, 0.0} 
\definecolor{color5}{rgb}{0.5, 0.0, 1.0} 
\addplot+[fill=color1] coordinates {(0,0)};
\addplot+[fill=color2] coordinates {(0,0)};
\addplot+[fill=color3] coordinates {(0,0)};
\addplot+[fill=color4] coordinates {(0,0)};
\legend{twitter7, com-Friendster, webbase, GAP-urand}
\end{axis}
\end{tikzpicture}

\begin{minipage}[b]{0.48\textwidth}
    \centering  
    \begin{tikzpicture}
     \begin{axis}[
    xlabel={Number of Processes},
    ylabel={Runtime(s)},
    align=center,
    ylabel style={yshift=-0.2em}, 
    legend style={at={(0.5,1.05)}, anchor=south, draw=none, legend columns=2, column sep=1ex},
    log basis x={2}, 
    log basis y={2}, 
    xmode=log, 
    ymode=log,
    xmin = 0.5,
    ymin = 1,
    ymax = 16384,
    xtick={1, 4, 16, 64, 256, 1024},
    ytick={1, 4, 16, 64, 256, 1024, 4096, 16384},
    xticklabel style={rotate=45, anchor=east, yshift=-0.5em}, 
    grid=both, 
    legend cell align={left},
    scale only axis,
    scale=0.55,
    width=\linewidth,
    height=\linewidth
]

\definecolor{color1}{rgb}{0.0, 0.0, 1.0} 
\definecolor{color2}{rgb}{1.0, 0.0, 0.0} 
\definecolor{color3}{rgb}{0.0, 1.0, 0.0} 
\definecolor{color4}{rgb}{1.0, 0.6, 0.0} 
\definecolor{color5}{rgb}{0.5, 0.0, 1.0} 

\addplot[solid, thick, color=color1, forget plot, mark=square*, mark options={scale=1}] table [x index=0, y index=1] {\scalingExtract}; 
\addplot[dashed, thick, color=color1, forget plot, mark=square*, mark options={scale=1}] table [x index=0, y index=6] {\scalingExtract}; 
\addlegendimage{only marks, mark=square*, mark options={scale=1, fill}, fill=color1}

\addplot[solid, thick, color=color2, forget plot, mark=triangle*, mark options={scale=1}] table [x index=0, y index=2] {\scalingExtract}; 
\addplot[dashed, thick, color=color2, forget plot, mark=triangle*, mark options={scale=1}] table [x index=0, y index=7] {\scalingExtract}; 
\addlegendimage{only marks, mark=triangle*, mark options={scale=1, fill}, fill=color2}

\addplot[solid, thick, color=color3, forget plot, mark=diamond*, mark options={scale=1}] table [x index=0, y index=3] {\scalingExtract}; 
\addplot[dashed, thick, color=color3, forget plot, mark=diamond*, mark options={scale=1}] table [x index=0, y index=8] {\scalingExtract}; 
\addlegendimage{only marks, mark=diamond*, mark options={scale=1, fill}, fill=color3}

\addplot[solid, thick, color=color4, forget plot, mark=oplus*, mark options={scale=1}] table [x index=0, y index=4] {\scalingExtract}; 
\addplot[dashed, thick, color=color4, forget plot, mark=oplus*, mark options={scale=1}] table [x index=0, y index=9] {\scalingExtract}; 
\addlegendimage{only marks, mark=oplus*, mark options={scale=1, fill}, fill=color4}
\end{axis}
\end{tikzpicture}
\subcaption{Submatrix extraction with 10\% of rows and columns.}\label{fig-scalability-extract-10}
\end{minipage}%
\begin{minipage}[b]{0.48\textwidth}
    \centering 
    \begin{tikzpicture}
    \begin{axis}[
    xlabel={Number of Processes},
    ylabel={Runtime(s)},
    align=center,
    ylabel style={yshift=-0.2em}, 
    legend style={at={(0.5,1.05)}, anchor=south, draw=none, legend columns=2, column sep=1ex},
    log basis x={2}, 
    log basis y={2}, 
    xmode=log, 
    ymode=log,
    xtick={1, 4, 16, 64, 256, 1024},
    xmin = 0.5,
    ymin = 1,
    ymax = 16384,
    ytick={1, 4, 16, 64, 256, 1024, 4096, 16384},
    xticklabel style={rotate=45, anchor=east, yshift=-0.5em}, 
    grid=both, 
    legend cell align={left},
    scale only axis,
    scale=0.55,
    width=\linewidth,
    height=\linewidth
]

\definecolor{color1}{rgb}{0.0, 0.0, 1.0} 
\definecolor{color2}{rgb}{1.0, 0.0, 0.0} 
\definecolor{color3}{rgb}{0.0, 1.0, 0.0} 
\definecolor{color4}{rgb}{1.0, 0.6, 0.0} 
\definecolor{color5}{rgb}{0.5, 0.0, 1.0} 

\addplot[solid, thick, color=color1, forget plot, mark=square*, mark options={scale=1}] table [x index=0, y index=1] {\scalingExtractt}; 
\addplot[dashed, thick, color=color1, forget plot, mark=square*, mark options={scale=1}] table [x index=0, y index=6] {\scalingExtractt}; 
\addlegendimage{only marks, mark=square*, mark options={scale=1, fill}, fill=color1}

\addplot[solid, thick, color=color2, forget plot, mark=triangle*, mark options={scale=1}] table [x index=0, y index=2] {\scalingExtractt}; 
\addplot[dashed, thick, color=color2, forget plot, mark=triangle*, mark options={scale=1}] table [x index=0, y index=7] {\scalingExtractt}; 
\addlegendimage{only marks, mark=triangle*, mark options={scale=1, fill}, fill=color2}

\addplot[solid, thick, color=color3, forget plot, mark=diamond*, mark options={scale=1}] table [x index=0, y index=3] {\scalingExtractt}; 
\addplot[dashed, thick, color=color3, forget plot, mark=diamond*, mark options={scale=1}] table [x index=0, y index=8] {\scalingExtractt}; 
\addlegendimage{only marks, mark=diamond*, mark options={scale=1, fill}, fill=color3}

\addplot[solid, thick, color=color4, forget plot, mark=oplus*, mark options={scale=1}] table [x index=0, y index=4] {\scalingExtractt}; 
\addplot[dashed, thick, color=color4, forget plot, mark=oplus*, mark options={scale=1}] table [x index=0, y index=9] {\scalingExtractt}; 
\addlegendimage{only marks, mark=oplus*, mark options={scale=1, fill}, fill=color4}

\end{axis}
    \end{tikzpicture}
\subcaption{Submatrix extraction with 50\% of rows and columns.}\label{fig-scalability-extract-50}
\end{minipage}%
\caption{Scalability of HipExtract (solid line) compared to CombBLAS extract (dashed line) when randomly extracting submatrices. 
Out-of-memory errors prevent the extract algorithms from running on larger matrices such as {\tt GAP-urand}, {\tt com-Friendster}, and {\tt webbase} on a single node (with 1 or 4 processes).  
}
\label{fig-scalability-extract}
\end{figure}
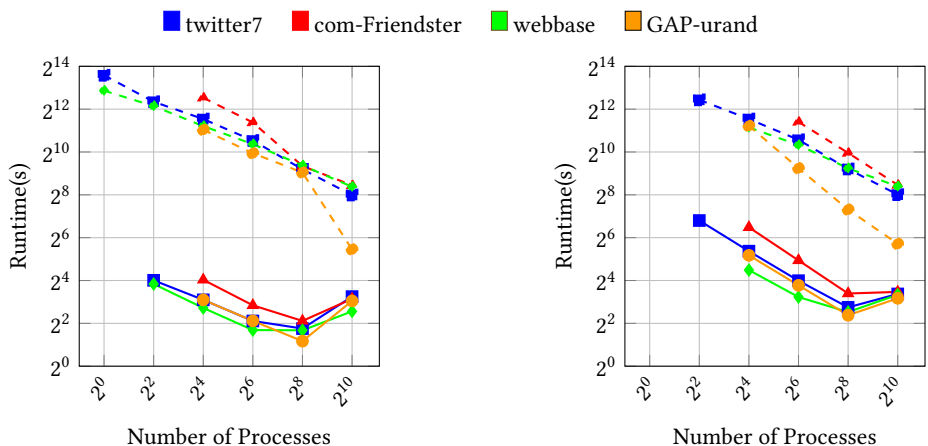

\subsection{Performance Evaluation of Submatrix Extraction}
We evaluate submatrix extraction in two scenarios: (a) extracting a random submatrix from a larger matrix, and (b) extracting communities or clusters from the adjacency matrix of a graph.
In both cases, we compare our algorithm, HipExtract, with the extract operation ({\tt SubsRef\_SR}) in CombBLAS. 

\paragraph{Extracting random submatrices.}
Extracting randomly selected submatrices is useful in applications such as sampling and sparsification. 
In this experiment, we randomly select 10\% or 50\% of the rows and columns of a matrix and extract the nonzeros corresponding to the selected indices. 
When the matrix represents a graph, this process corresponds to extracting a random subgraph.

Figure~\ref{fig-scalability-extract} shows the scalability of HipExtract and CombBLAS's extract for randomly extracting 10\% and 50\% submatrices.
In both scenarios, HipExtract runs about an order of magnitude faster than CombBLAS.
Both algorithms show good scalability up to 256 processes, with HipExtract consistently outperforming CombBLAS.
Figure~\ref{fig-scalability-extract} also reveals that HipExtract stops scaling after 256 processes.
This occurs because smaller extracted subgraphs limit the benefits of parallelism at higher process counts.
For the same input matrices, HipExtract runs faster when extracting 10\% rather than 50\% of the matrix. 
In contrast, CombBLAS shows little change, with runtimes remaining nearly constant. This is because CombBLAS’s extract is dominated by the input matrix size, whereas HipExtract’s runtime scales with the size of the extracted submatrix.

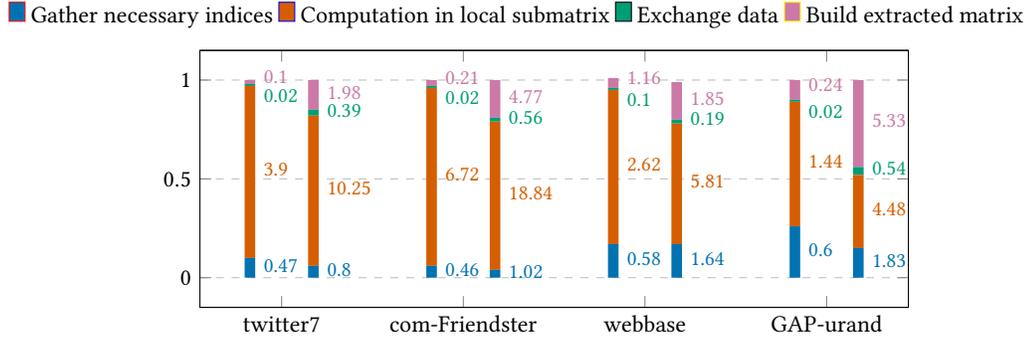
\begin{figure}[!t]
\centering
\begin{tikzpicture}
\begin{axis}[
    height=5cm,
    width=11cm,
    ybar stacked,
    bar width=4pt,
    enlargelimits=0.15,
    ymin=0, ymax=1,
    xtick=data,
    xticklabels={twitter7, com-Friendster, webbase, GAP-urand},
    legend style={at={(0.45,1.05)}, anchor=south, draw=none, legend columns=4},
    ymajorgrids=true,
    grid style=dashed,
    bar shift={-12pt},
    cycle list name=color list,
    nodes near coords,
    point meta=explicit symbolic,
    nodes near coords style={
        font=\small, anchor=west, xshift=2pt, yshift=1pt
    },
]
\addplot+[fill=oiBlue, draw=none, every node near coord/.append style={oiBlue }] coordinates {(0,0.10)[0.47] (1,0.06)[0.46] (2,0.17)[0.58] (3,0.26)[0.6]};
\addlegendentry{Gather necessary indices}
\addplot+[fill=oiRed, draw=none, every node near coord/.append style={oiRed}]  coordinates {(0,0.87)[3.9] (1,0.90)[6.72] (2,0.78)[2.62] (3,0.63)[1.44]};
\addlegendentry{Computation in local submatrix}
\addplot+[fill=oiGreen, draw=none, every node near coord/.append style={oiGreen,  yshift=-5pt}]coordinates {(0,0.01)[0.02] (1,0.01)[0.02] (2,0.01)[0.1] (3,0.01)[0.02]};
\addlegendentry{Exchange data}
\addplot+[fill=oiPink, draw=none, every node near coord/.append style={oiPink, yshift=1pt}]coordinates {(0,0.02)[0.1] (1,0.03)[0.21] (2,0.05)[1.16] (3,0.1)[0.24]};
\addlegendentry{Build extracted matrix}
\end{axis}

\begin{axis}[
    height=5cm,
    width=11cm,
    ybar stacked,
    bar width=4pt,
    enlargelimits=0.15,
    ymin=0, ymax=1,
    hide y axis,
    xtick=\empty,
    bar shift={+12pt},
    cycle list name=color list,
    nodes near coords,
    point meta=explicit symbolic,
    nodes near coords style={
        font=\small, anchor=west, xshift=2pt, yshift=1pt
    },
]
			
\addplot+[fill=oiBlue, draw=none, every node near coord/.append style={oiBlue }] coordinates {(0,0.06)[0.8] (1,0.04)[1.02] (2,0.17)[1.64] (3,0.15)[1.83]};
\addplot+[fill=oiRed, draw=none, every node near coord/.append style={oiRed }]  coordinates {(0,0.76)[10.25] (1,0.75)[18.84] (2,0.61)[5.81] (3,0.37)[4.48]};
\addplot+[fill=oiGreen, draw=none, every node near coord/.append style={oiGreen }]coordinates {(0,0.03)[0.39] (1,0.02)[0.56] (2,0.02)[0.19] (3,0.04)[0.54]};
\addplot+[fill=oiPink, draw=none, every node near coord/.append style={oiPink}]coordinates {(0,0.15)[1.98] (1,0.19)[4.77] (2,0.19)[1.85] (3,0.44)[5.33]};
\end{axis}

\end{tikzpicture}
\caption{Breakdown of runtime for extracting 10\% (left bar) and 50\% (right bar) of randomly selected nonzeros using HipExtract. 
}
\label{fig-breakdown-extract}
\end{figure}

To analyze the runtime of each step of HipExtract, we present a breakdown of the runtime in Figure ~\ref{fig-breakdown-extract}.
The results indicate that the local computation step, {\em Identifying extracting elements}, dominates the runtime of HipExtract. 
Specifically, for all matrices in Figure~\ref{fig-breakdown-extract}, when extracting a smaller submatrix with 10\% of the rows and columns, the runtime is dominated by the time spent on \emph{identifying the elements to extract}.

\begin{figure}[!t]
\begin{minipage}{0.4\textwidth} 
\centering
\subcaption{\small Columns "Nodes" and "NNZ" correspond to the input graph, while "Agg. Nodes" and "Agg. NNZ" show total nodes and nonzeros extracted from the 10 largest clusters.}
\label{fig-ext-clust-scalability-table}
\renewcommand{\arraystretch}{1.2} 
\begin{tabular}{|c|c|c|c|c|}
    \hline
    Dataset & Nodes & NNZ & \makecell{Agg.\\Nodes} & \makecell{Agg.\\NNZ} \\
    \hline
    Virus   & 219K & 4.5M & 2.5K & 512K\\
    Eukarya  & 3.2M & 359M & 78K & 22K\\
    Archaea  & 5.5M & 148M & 44K & 45K\\
    \hline
\end{tabular}
\end{minipage}%
\hfill 
\begin{minipage}{0.58\textwidth} 
\centering
\begin{tikzpicture}
    \begin{axis}[
        xlabel={Number of Processes},
        ylabel={Runtime(s)},
        align=center,
        ylabel style={yshift=-0.2em}, 
        ytick={1/8, 1, 8, 64, 512, 4096, 32768 },
        legend style={at={(0.45,1.05)}, anchor=south, draw=none, legend columns=3, column sep=1ex},
        log basis x={2}, 
        log basis y={2}, 
        xmode=log, 
        ymode=log,
        xtick=data, 
        xticklabel style={rotate=45, anchor=east, yshift=-0.5em}, 
        grid=both, 
        legend cell align={left},
        scale only axis,
        scale=0.45,
        width=\linewidth,
        height=\linewidth
    ]
    \definecolor{color1}{rgb}{0.0, 0.0, 1.0} 
    \definecolor{color2}{rgb}{1.0, 0.0, 0.0} 
    \definecolor{color3}{rgb}{0.0, 1.0, 0.0} 
    
    \addplot[solid, thick, color=color1, forget plot, mark=square*, mark options={scale=1}] table [x index=0, y index=1] {\scalingExtClust}; 
    \addplot[dashed, thick, color=color1, forget plot, mark=square*, mark options={scale=1}] table [x index=0, y index=4] {\scalingExtClust}; 
    \addlegendimage{only marks, mark=square*, mark options={scale=1, fill}, fill=color1}
    \addlegendentry{Virus}

    \addplot[solid, thick, color=color2, forget plot, mark=diamond*, mark options={scale=1}] table [x index=0, y index=2] {\scalingExtClust}; 
    \addplot[dashed, thick, color=color2, forget plot, mark=diamond*, mark options={scale=1}] table [x index=0, y index=5] {\scalingExtClust}; 
    \addlegendimage{only marks, mark=diamond*, mark options={scale=1, fill}, fill=color2}
    \addlegendentry{Eukarya}

    \addplot[solid, thick, color=color3, forget plot, mark=oplus*, mark options={scale=1}] table [x index=0, y index=3] {\scalingExtClust}; 
    \addplot[dashed, thick, color=color3, forget plot, mark=oplus*, mark options={scale=1}] table [x index=0, y index=6] {\scalingExtClust}; 
    \addlegendimage{only marks, mark=oplus*, mark options={scale=1, fill}, fill=color3}
    \addlegendentry{Archaea}
    \end{axis}
\end{tikzpicture}
\subcaption{Scalability of our extract algorithm compared to CombBLAS.}
\label{fig-ext-clust-scalability-fig}
\end{minipage}
\caption{Performance of HipExtract compared to CombBLAS when extracting the 10 largest clusters in the corresponding datasets. 
}
\label{fig-ext-clust-scalability}
\end{figure}
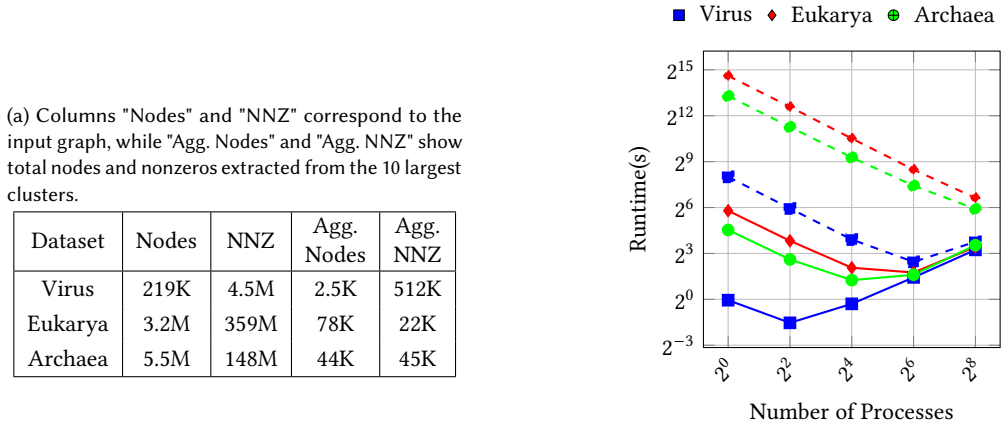

\paragraph{Extracting clusters from a graph.}
In this experiment, we extract subgraphs representing clusters or communities within biological networks. 
 
We use three protein-similarity networks: {\tt Virus}, {\tt Eukarya}, and {\tt Archaea}, as described in Table ~\ref{fig-ext-clust-scalability}(a). 
These graphs contain thousands of clusters, where each cluster corresponds to a family of proteins with similar sequences and functions. 
We first applied the Markov Clustering Algorithm (MCL)~\cite{azad2018hipmcl,selvitopi2020optimizing} to identify communities and then extracted the 10 largest clusters from each graph.
Figure~\ref{fig-ext-clust-scalability}(b) shows that HipExtract is an order of magnitude faster than CombBLAS’s Extract for the larger {\tt Eukarya} and {\tt Archaea} graphs. For the smaller {\tt Virus} graph, however, its scalability is limited at higher process counts because the extracted submatrices become very small when distributed across many processes (see the last column of Table~\ref{fig-ext-clust-scalability}(a)).

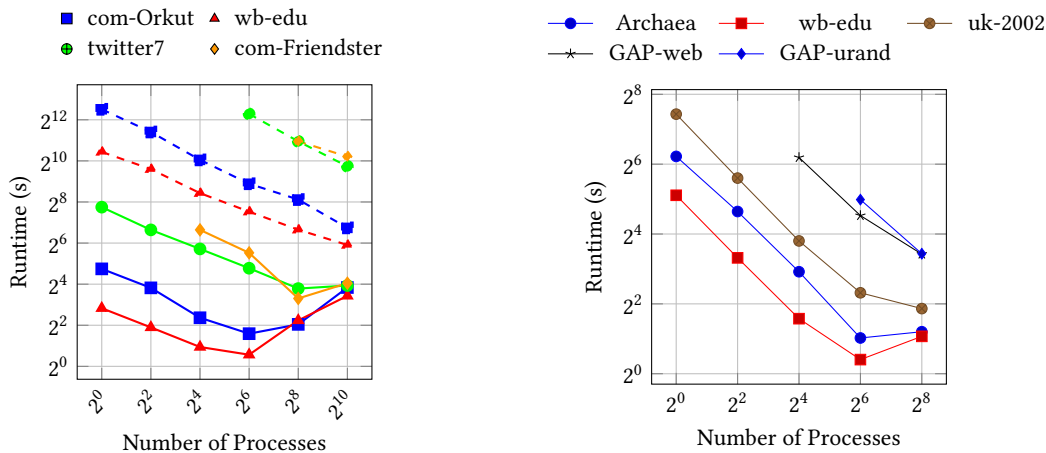
\begin{figure}[htbp]
\centering
\begin{minipage}{0.52\textwidth}
\centering
\begin{tikzpicture}
    \begin{axis}[
        xlabel={Number of Processes},
        ylabel={Runtime (s)},
        ylabel style={yshift=-0.3em},
        xlabel style={yshift=0.3em},
        log basis x=2,
        log basis y=2,
        width=5.5cm,
        height=5.5cm,
        xmode=log,
        ymode=log,
        xtick=data,
        ytick={1, 4, 16, 64, 256, 1024, 4096},
        xticklabel style={rotate=45, anchor=east, yshift=-0.5em},
        grid=both,
        legend cell align={left},
        legend style={at={(0.5,1.05)}, anchor=south, draw=none, legend columns=2, column sep=1ex}
    ]

    \definecolor{color1}{rgb}{0.0, 0.0, 1.0}
    \definecolor{color2}{rgb}{1.0, 0.0, 0.0}
    \definecolor{color3}{rgb}{0.0, 1.0, 0.0}
    \definecolor{color4}{rgb}{1.0, 0.6, 0.0}

    \addplot[solid, thick, color=color1, forget plot, mark=square*, mark options={scale=1}] table [x index=0, y index=1] {\scalingAsgn}; 
    \addplot[dashed, thick, color=color1, forget plot, mark=square*, mark options={scale=1}] table [x index=0, y index=6] {\scalingAsgn}; 
    \addlegendimage{only marks, mark=square*, mark options={scale=1, fill}, fill=color1}
    \addlegendentry{com-Orkut}

    \addplot[solid, thick, color=color2, forget plot, mark=triangle*, mark options={scale=1}] table [x index=0, y index=2] {\scalingAsgn}; 
    \addplot[dashed, thick, color=color2, forget plot, mark=triangle*, mark options={scale=1}] table [x index=0, y index=7] {\scalingAsgn}; 
    \addlegendimage{only marks, mark=triangle*, mark options={scale=1, fill}, fill=color2}
    \addlegendentry{wb-edu}

    \addplot[solid, thick, color=color3, forget plot, mark=oplus*, mark options={scale=1}] table [x index=0, y index=3] {\scalingAsgn}; 
    \addplot[dashed, thick, color=color3, forget plot, mark=oplus*, mark options={scale=1}] table [x index=0, y index=8] {\scalingAsgn}; 
    \addlegendimage{only marks, mark=oplus*, mark options={scale=1, fill}, fill=color3}
    \addlegendentry{twitter7}

    \addplot[solid, thick, color=color4, forget plot, mark=diamond*, mark options={scale=1}] table [x index=0, y index=4] {\scalingAsgn}; 
    \addplot[dashed, thick, color=color4, forget plot, mark=diamond*, mark options={scale=1}] table [x index=0, y index=9] {\scalingAsgn}; 
    \addlegendimage{only marks, mark=diamond*, mark options={scale=1, fill}, fill=color4}
    \addlegendentry{com-Friendster}
    
    \end{axis}
\end{tikzpicture}
\subcaption{Scalability of HipAssign (solid line) compared to CombBLAS assign (dashed line) when assignment is performed without permutation.}
\label{fig-asgn-scalability}
\end{minipage}
\hfill
\begin{minipage}{0.44\textwidth}
\centering
\begin{tikzpicture}
    \begin{axis}[
        xlabel={Number of Processes},
        ylabel={Runtime (s)},
        width=5.5cm,
        height=5.5cm,
        %
        log basis x=2,
        log basis y=2,
        xmode=log,
        ymode=log,
        xtick={1,4,16,64,256},
        ytick={1, 4, 16, 64, 256, 1024, 4096},
        grid=both,
        samples=5,
        legend style={at={(0.5,1.05)}, anchor=south, draw=none, legend columns=3, column sep=1ex},
        legend entries={
            Archaea, 
            wb-edu, 
            uk-2002, 
            GAP-web, 
            GAP-urand
        },
    ]
    \addplot coordinates {(1,74.61) (4,25) (16,7.568) (64,2.033) (256,2.302)};
    \addplot coordinates {(1,34.51) (4,9.97) (16,2.982) (64,1.324) (256,2.1)};
    \addplot coordinates {(1,172.59) (4,48.527) (16,13.946) (64,4.984) (256,3.643)};
    \addplot coordinates {(16,72.999) (64,23.068) (256,10.73)};
    \addplot coordinates {(64,31.66) (256,10.82)};
    \end{axis}
\end{tikzpicture}
\subcaption{Scalability of HipAssign when assignment is followed by permutation for load balance.}
\label{fig-update-scalability}
\end{minipage}
\caption{Scalability of HipAssign when assigning submatrices in the streaming batch update experiment.}
\end{figure}

\subsection{Performance Evaluation of Submatrix Assign}
We evaluate the performance of HipAssign by updating a graph with subgraphs.
To capture a streaming graph application, we partition an input graph into five parts: one containing 80\% of the original edges and four smaller subgraphs with 5\% of the edges each. 
The assign operation is then performed in four steps, where the adjacency matrices of the smaller subgraphs are incrementally merged into the larger graph. 
The final result of these four assignments is the original input graph with its full set of edges. The runtimes of these four assign operations, measured for both our algorithm and CombBLAS, are reported in Figure~\ref{fig-asgn-scalability}.

The speedup of our assign algorithm to CombBLAS's assign is 182.4 on average, with stdev=29.2, but it highly depends on the structure of the graph and the assignment indexing vectors. 
In particular, CombBLAS failed to complete the assignment for larger graphs such as {\tt Twitter7} and {\tt com-Friendster} on fewer nodes due to out-of-memory errors. 
We observe that HipAssign stops scaling at 64 processes for smaller graphs and at 256 processes for larger graphs. 
This behavior arises because HipAssign’s runtime is proportional to the size of the subgraph being assigned. 
As a result, scalability stalls when a small subgraph is distributed across many processes.

Batch updates of graphs in real applications generally result in load imbalance, which can affect the performance of subsequent distributed analysis algorithms.
Permuting graphs during the update process can be a solution to this issue.
Using our algorithm, batch update and permutation can be fused to reduce the overhead of performing a separate permutation step \cite{hassani2024batch}.
Figure~\ref{fig-update-scalability} illustrates the performance and scalability of the fused update with the permute algorithm, which is scaling across different datasets.
In this experiment, we update each graph using a randomly generated \erdosrenyi~ graph in a single step.
Runtimes in some cases were omitted due to memory limitations. 
Permuting the matrix during the update preserves the load balance between processes without incurring additional computational cost.
Compared to performing assign and permute separately using CombBLAS, our method is at least 25$\times$ faster while preserving the load balance.

\subsection{Fusing Operations in an Incremental Graph Application}
A key benefit of the IEB approach is its ability to fuse multiple operations with little overhead. To illustrate this benefit, we consider incremental graph updates where new subgraphs are added to an existing graph. 
An example arises in protein sequence databases such as the IMG/M database~\cite{chen2019img}.
In some applications, such databases are modeled as a graph where vertices denote proteins, and edges connect proteins with similarity above a threshold. As new sequences arrive, they form subgraphs that are integrated into the existing graph.

Incremental graph updates correspond to repeated assignments of submatrices. 
In a distributed-memory setting, this may lead to load imbalance, requiring periodic matrix permutations. 
While assignment and permutation can be invoked sequentially (e.g., HipAssign followed by random HipPerm), doing so adds runtime overhead. 
By modifying the Identify step, we developed a fused HipAssign+HipPerm operation that performs both tasks simultaneously. Its performance is analyzed in this section.

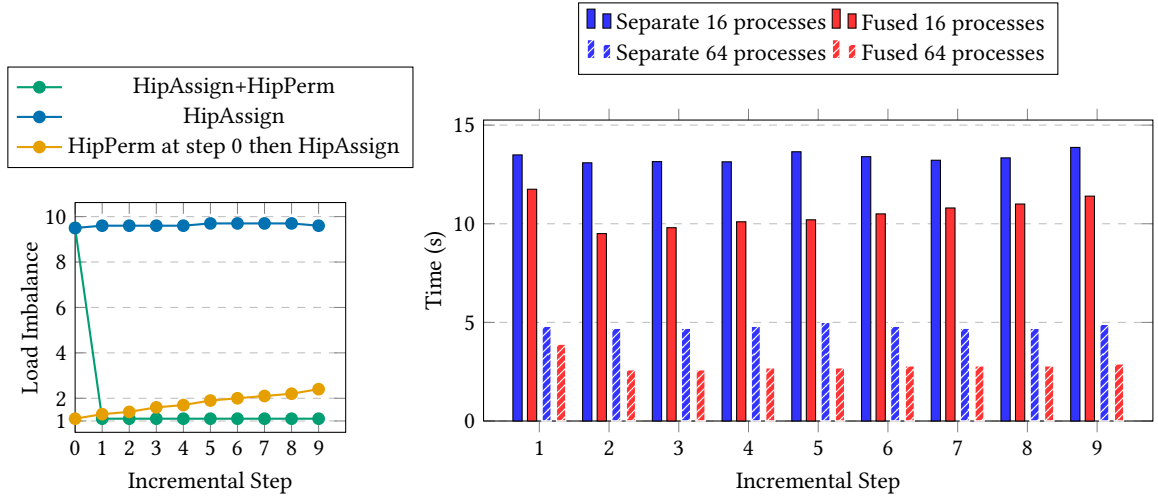
\begin{figure*}[htbp]
\centering
\begin{minipage}[b]{0.34\textwidth}
\centering 
\begin{tikzpicture}
\begin{axis}[
    width=\linewidth,
    height=0.85\linewidth,
    scale=0.7,
    scale only axis,
    xlabel={Incremental Step},
    ylabel={Load Imbalance},
    ylabel style={yshift=-0.4em}, 
    xmin=0, xmax=10,
    xtick={0,1,2,3,4,5,6,7,8,9},
    ymin=0.5,
    ytick={1,2,4,6,8,10},
    legend style={at={(0.5,1.15)}, anchor=south,legend columns=1},
    ymajorgrids=true,
    grid style=dashed
]
\addplot[color=oiGreen, mark=*, thick] coordinates { (0,9.5) (1,1.1) (2,1.1) (3,1.1) (4,1.1) (5,1.1) (6,1.1) (7,1.1) (8,1.1) (9,1.1)};
\addplot[color=oiBlue, mark=*, thick] coordinates { (0,9.5) (1,9.6) (2,9.6) (3,9.6) (4,9.6) (5,9.7) (6,9.7) (7,9.7) (8,9.7) (9,9.6)};
\addplot[color=oiOrange, mark=*, thick] coordinates { (0,1.1) (1,1.3) (2,1.4) (3,1.6) (4,1.7) (5,1.9) (6,2) (7,2.1) (8,2.2) (9,2.4)};
\legend{
HipAssign+HipPerm,
HipAssign,
HipPerm at step 0 then HipAssign,
}
\end{axis}
\end{tikzpicture}
\subcaption{Load imbalance of the matrix across 64 processes after each incremental update. Step 0 denotes the state before any updates.}
\label{fig-incremental-uk-1}
\end{minipage}
\hfill
\begin{minipage}[b]{0.64\textwidth}
\centering 
\begin{tikzpicture}
\begin{axis}[
    ybar,
    bar width=3.5pt,
    width=\linewidth,
    height=0.45\linewidth,
    scale=0.92,
    scale only axis,
    ylabel={Time (s)},
    ylabel style={yshift=-0.4em}, 
    xlabel={Incremental Step},
    symbolic x coords={1,2,3,4,5,6,7,8,9},
    xtick=data,
    ymin=0,
    ytick={0, 5, 10, 15, 20, 25},
    legend style={at={(0.5,1.15)}, anchor=south,legend columns=2},
    ymajorgrids=true,
    grid style=dashed,
]
\addplot[fill=blue!80] coordinates {(1,13.49) (2,13.09) (3,13.15) (4,13.14) (5,13.65) (6,13.4) (7,13.22) (8,13.34) (9,13.87)}; 
\addplot[fill=red!80] coordinates {(1,11.75) (2,9.5) (3,9.8) (4,10.1) (5,10.2) (6,10.5) (7,10.8) (8,11) (9,11.4)}; 
\addplot[white, fill=blue!80, postaction={pattern=north east lines, pattern color=white}] coordinates {(1,4.8) (2, 4.7) (3,4.7) (4,4.8) (5,5) (6,4.8) (7,4.7) (8,4.7) (9,4.9)}; 
\addplot[white, fill=red!80, postaction={pattern=north east lines, pattern color=white}] coordinates {(1,3.9) (2,2.6) (3,2.6) (4,2.7) (5,2.7) (6,2.8) (7,2.8) (8,2.8) (9,2.9)}; 
\legend{
 Separate 16 processes,
 Fused 16 processes,
Separate 64 processes,
 Fused 64 processes
}
\end{axis}
\end{tikzpicture}
\subcaption{Runtime comparison of (i) separate HipAssign+HipPerm and (ii) fused HipAssign+HipPerm at each step of the incremental update experiment on 16 and 64 processes.}
\label{fig-incremental-uk-2}
\end{minipage}
\caption{Incremental update of {\tt uk-2002}, where nine small submatrices (each $\sim$2\% of edges) are assigned to a large submatrix containing 80\% of edges. Experiments were conducted on the Grace supercomputer with 16 and 64 processes, each using 12 threads.
}
\label{fig-incremental-uk}
\end{figure*}

\begin{figure*}[htbp]
\centering
\begin{minipage}[b]{0.34\textwidth}
\centering 
\begin{tikzpicture}
\begin{axis}[
    width=\linewidth,
    height=0.75\linewidth,
    scale=0.7,
    scale only axis,
    xlabel={Incremental Step},
    ylabel={Load Imbalance},
    ylabel style={yshift=-0.4em}, 
    xmin=0, xmax=10,
    xtick={0,1,2,3,4,5,6,7,8,9},
    ymin=0.95,
    ymax = 1.25,
    ytick={1,1.14,1.8},
    legend style={at={(0.5,1.15)}, anchor=south,legend columns=1},
    ymajorgrids=true,
    grid style=dashed
]
\addplot[color=oiGreen , mark=*, thick] coordinates { (0,1.14) (1,1.0) (2,1.0) (3,1.0) (4,1.0) (5,1.0) (6,1.0) (7,1.0) (8,1.0) (9,1.0)};
\addplot[color=oiBlue , mark=*, thick] coordinates { (0,1.14) (1,1.14) (2,1.15) (3,1.14) (4,1.14) (5,1.14) (6,1.15) (7,1.13) (8,1.14) (9,1.14)};
\legend{
HipAssign+HipPerm,
HipAssign,
}
\end{axis}
\end{tikzpicture}
\subcaption{Load imbalance of matrix after each step of incremental update ran on 64 processes. 
Step 0 represent the matrix before any increments.}
\label{fig-incremental-metaclust-1}
\end{minipage}
\hfill
\begin{minipage}[b]{0.64\textwidth}
\centering 

\begin{tikzpicture}
\begin{axis}[
    ybar,
    bar width=4.5pt,
    width=\linewidth,
    height=0.45\linewidth,
    scale=0.9,
    scale only axis,
    ylabel={Time (s)},
    ylabel style={yshift=-0.4em}, 
    xlabel={Incremental Step},
    symbolic x coords={1,2,3,4,5,6,7,8,9},
    xtick=data,
    ymin=45,
    ytick={45, 50, 55, 60, 65, 70, 75, 80},
    legend style={at={(0.5,1.15)}, anchor=south,legend columns=2},
    ymajorgrids=true,
    grid style=dashed,
]
\addplot[fill=blue!80, draw=none] coordinates {(1,69.75) (2, 71.52) (3,70.46) (4,78.99) (5,74.45) (6,73.32) (7,74.53) (8,76.41) (9,77.24)}; 
\addplot[fill=red!80, draw=none] coordinates {(1,66.79) (2,71.70) (3,68.92) (4,75.31) (5,70.00) (6,71.71) (7,71.26) (8,71.57) (9,72.82)}; 
\legend{
 Separate 256 processes,
 Fused 256 processes
}
\end{axis}
\end{tikzpicture}

\subcaption{Runtime comparison of applying (i) Seperate HipAssign+HipPerm and (ii) Fused HipAssign+HipPerm at each step of incremental update experiment.
Runtime is shown for 256 and 1024 processes.}
\label{fig-incremental-metaclust-2}
\end{minipage}
\caption{
Incremental update of {\tt Metaclust50}, where nine small submatrices (each $\sim$2\% of edges) are assigned to a large submatrix containing 80\% of edges. Experiments were conducted on the Perlmutter supercomputer with 1024 and 256 processes, each using 16 threads.}
\label{fig-incremental-metaclust}
\end{figure*}
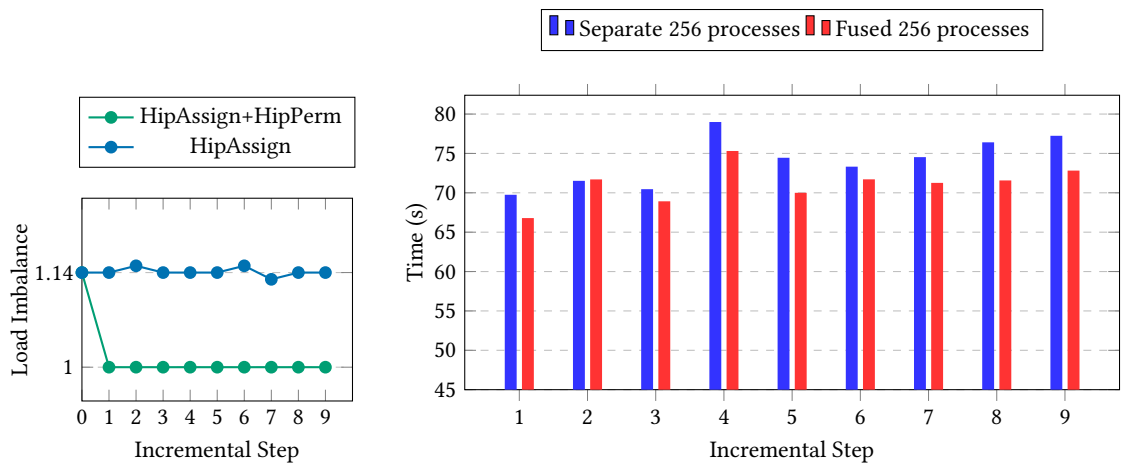

We compare the fused HipAssign+HipPerm with their separate invocation on two graphs, {\tt uk-2002} and {\tt Metaclust50}. 
Each matrix is partitioned into ten non-overlapping submatrices: one large submatrix containing about 80\% of the nonzeros, and nine smaller ones each containing about 2\%. Submatrices are formed by randomly selecting subsets of rows and columns using HipExtract. 
The fused operation is then applied incrementally in nine steps, where each small submatrix is assigned to the large one. 
This setup simulates incremental matrix updates while maintaining load balance in a distributed environment.
Figure~\ref{fig-incremental-uk} and Figure~\ref{fig-incremental-metaclust} show the results for {\tt uk-2002} and {\tt Metaclust50}, respectively.

Figure~\ref{fig-incremental-uk-1} shows that without permutation during incremental updates (orange line), load imbalance remains high and gradually increases. 
Permuting once at step 0 (blue line) reduces the imbalance initially, but it grows over time. 
In contrast, the fused HipAssign+HipPerm (yellow line) consistently maintains load imbalance near 1.0. On average across nine steps. 
The fused method is about 20\% faster than separate HipAssign+HipPerm with 16 processes and 38\% faster with 64 processes, where HipAssign takes 18\% of the time.

Figure~\ref{fig-incremental-metaclust-1} shows that the load imbalance remains at $1.0$ when fused HipAssign+HipPerm is applied(yellow line) and remains around $1.14$ when HipPerm is not applied(orange line).
Given the balanced nature of this matrix, the fused method is about 4\% faster than HipAssign+HipPerm and HipAssign is taking 9\% of the total runtime for HipAssign+HipPerm. 
This indicates that HipAssign is a relatively small fraction of the overall cost. 
Since the cost of HipAssign can be amortized in applications where permutation is required, the fused method hides the assignment cost and makes it negligible in practice.
For matrices where updates naturally preserve balance, permutations can be applied less frequently.

\section{Conclusions}
Matrix permutation, extraction, and assignment are fundamental operations in sparse matrix algorithms and libraries. 
Their performance is especially critical in distributed-memory settings, where inefficient design can lead to significant communication overheads. 
Despite their ubiquity, the literature on parallel algorithms for these operations remains limited.

We showed that direct implementations of permutation, extraction, and assignment using the identify–exchange–build (IEB) approach outperform SpGEMM-based methods in distributed memory. 
For example, our experiments demonstrated that HipPerm and the permutation routine in PETSc achieve better performance than the SpGEMM-based implementation in CombBLAS. 
The performance gap arises because CombBLAS’s SpGEMM treats both operands symmetrically, whereas our algorithms communicate only the required data, leading to significant reductions in both communication and overall runtime.

Efficient and scalable implementations also require highly optimized multithreaded local kernels. For fine-grained matrix access and updates, minimizing thread synchronization is crucial to achieving high performance.
This paper presented shared-memory parallel algorithms that eliminate synchronization, enabling linear thread scalability within a process.

In many applications, these operations must be combined. 
The IEB-style framework naturally supports such fusion, enabling multiple operations to be executed together with reduced communication. 
Although we illustrated this in the context of a streaming graph application, the approach also generalizes to other scenarios.

In summary, this work provides scalable parallel algorithms and software for matrix permutation, extraction, and assignment. These implementations offer practical building blocks that can be integrated into existing sparse matrix libraries to enhance performance across a wide range of applications.









\section*{Acknowledgments}
This research was funded in part by DOE grants DE-SC0022098 and DE-SC0023349 and by NSF grants PPoSS CCF 2316233 and OAC-2339607.




\bibliographystyle{ACM-Reference-Format}
\bibliography{ref}

\appendix
\section{Appendix}
\subsection{Multithreaded algorithm for local computations in HipExtract}
Algorithm~\ref{algo:threadComputation-extract} outlines the multithreaded algorithm for extracting entries from the local matrix and populating the $\SendBuffer$ array.

\begin{algorithm}[!h]
\caption{Preparing Send Data for HipExtract at process $P_{i,j}$}
\label{algo:threadComputation-extract}
\textbf{Input and Output:} Local matrix $\mat{A}_{r,c}$, local permutation vectors $\vect{SpPvec\_loc }$ and $\vect{SpQvec\_loc}$. Output: The \SendBuffer ~ array ready for communication.
\begin{algorithmic}[1]
\Procedure{ExtractPrepareSendBuffer}{$\mat{A}_{r,c}, \vect{SpPvec\_loc}, \vect{SpQvec\_loc}$}
\State Logically partition $\vect{SpQvec\_loc}$ into $\nthreads$ equal parts
\State Partition $\mat{A}_{r,c}$ by column indices based on $\vect{SpQvec\_loc}$ partitioning
\State $\texttt{nnzCounter} \gets \text{zeros}(\texttt{nthreads} \times p)$ \Comment{Initialize a vector with zero values}
\For {$t \gets 1$ to \nthreads ~  \textbf{in parallel}} \Comment{OpenMP thread parallel}
     \State ($\Tilde{\mat{A}}_{r,c}$, $\Tilde{\vect{SpQvec\_loc}}$) $\gets$ the $t$-th partition of ($\mat{A}$, $\vect{SpQvec\_loc}$)
    \For{Each $(i,j)$ in $\vect{Indices}$ of ($\vect{SpPvec\_loc}$, $\Tilde{\vect{SpQvec\_loc}}$)}
    \If{Binary search of $(i)$ in nonzero row indices of $\Tilde{\mat{A}}_{r,c}[:,j]$ is true}
     \State \texttt{extElement} $\gets$ $\Tilde{\mat{A}}_{r,c}[i,j]$
    \State \texttt{DestP}$ \gets$ Compute the process that owns \texttt{extElement} in $\mat{B}$
    \State $\texttt{nnzCounter}[t,\texttt{DestP}]\gets \texttt{nnzCounter}[t,\texttt{DestP}] + 1$ 
    \EndIf
    \EndFor
\EndFor

\For {$t \gets 1$ to \nthreads ~  \textbf{in parallel}} \Comment{OpenMP thread parallel}
     \State ($\Tilde{\mat{A}}_{r,c}$, $\Tilde{\vect{SpQvec\_loc}}$) $\gets$ the $t$-th partition of ($\mat{A}$, $\vect{SpQvec\_loc}$)
      \For{Each $(i,j)$ in $\vect{Indices}$ of ($\vect{SpPvec\_loc}$, $\Tilde{\vect{SpQvec\_loc}}$)}
      \If{Binary search of $(i)$ in nonzero row indices of $\Tilde{\mat{A}}_{r,c}[:,j]$ is true}
    \State \texttt{extElement} $\gets$ $\Tilde{\mat{A}}_{r,c}[i,j]$
    \State \texttt{DestP} $\gets$ Compute the process that owns \texttt{extElement} in $\mat{B}$
        \State $(lrow, lcol) \gets $ local row and column index of \texttt{extElement} in \texttt{DestP}
        \State $index = \Call{SendBufferIndex}{\texttt{DestP}, \texttt{nnzCounter}}$ 
        \State $\SendBuffer[index] \gets (lrow, lcol,$\texttt{extElement}$)$
    \EndIf
    \EndFor
\EndFor
\State \Return \SendBuffer
\EndProcedure
\end{algorithmic}
\end{algorithm}

\end{document}